\documentclass[aps]{revtex4}
\def\llsymbol#1{\@llsymbol{\@nameuse{c@#1}}}
\def\@llsymbol#1{\ifcase#1\or {}\or {'}\or {''}\or {'''}\or
Â  Â {''''}\or {'''''}\or Â \else\@ctrerr\fi\relaz}
\newcounter{contador}
\newcommand{\letra}{
Â  Â \stepcounter{equation}
Â  Â \setcounter{contador}{\value{equation}}
Â  Â \setcounter{equation}{0}
Â  Â \renewcommand{\theequation}{\thecontador\alph{equation}}}
\newcommand{\antiletra}{
Â  Â \renewcommand{\theequation}{\arabic{equation}}
Â  Â \setcounter{equation}{\value{contador}}}

\usepackage{color}
\usepackage{indentfirst}
\usepackage{eucal}
\usepackage{amsmath}
\usepackage{amsfonts}
\usepackage{amssymb}
\usepackage{mathrsfs}
\begin{document}
\title{
Transformations of Heun's equation and its integral relations
\\{\small (To appear in J. Phys. A)}}
\author{ L\'ea Jaccoud El-Jaick}
\email{leajj@cbpf.br}
\author{ Bartolomeu D. B. Figueiredo}
\email{barto@cbpf.br}
\affiliation{Centro Brasileiro de Pesquisas F\'{\i}sicas (CBPF),\\
Rua Dr. Xavier Sigaud, 150, CEP 22290-180, Rio de Janeiro, RJ, Brasil
}

\begin{abstract}
\noindent
Using the transformation theory
for the Heun equation, we find substitutions of variables which
preserve the form of the equation for
the kernels of integral relations among solutions of the
Heun equation. These transformations lead to new kernels
for the Heun equation,
given by single hypergeometric
functions (Lambe-Ward-type kernels) and by products of two
hypergeometric functions (Erd\'elyi-type). Such kernels,
by a limiting process, also afford new
kernels for the confluent Heun equation.
\end{abstract}
\maketitle
%
%
%
%
\section*{1. Introductory remarks}
The group of transformations of
variables which changes Heun's equation
into another version of itself was initially
established by the very Heun
in 1889 \cite{heun} and fully accounted in 2007 
 by Maier \cite{maier}, who
tabulated the 192 substitutions in detail by
writing explicitly the transformations of
each parameter and variable of the equation.
Firstly, we show that it is possible to
construct a similar table for the kernels of
integral relations among solutions of the equation.

In the second place, we show that
some of these transformations generate new kernels
given by hypergeometric functions when applied
to the kernels found  by Lambe and
Ward in 1934 \cite{lambe}, and new kernels
in terms of products of hypergeometric functions
when applied to the kernels found by Erd\'elyi in 1942
\cite{erdelyi5}. Finally, by means of a limiting
procedure we get new kernels
also for the confluent Heun equation (CHE) .

The transformations of the Heun equation
and its integral relations suppose the usual algebraic
form for the equation \cite{heun,maier,arscott2},
namely,
\begin{eqnarray}\label{heun}
\frac{d^{2}H}{dx^{2}}+\left[\frac{\gamma}{x}+
\frac{\delta}{x-1}+\frac{\epsilon}{x-a}\right]\frac{dH}{dx}+
\left[ \frac{\alpha \beta x-q}{x(x-1)(x-a)}\right] H=0,
\quad [\epsilon=\alpha+\beta+1-\gamma-\delta]
\end{eqnarray}
where $a\in \mathbb{C}\setminus\{0,1\}$
and $x=0,1,a,\infty$ are regular singular points with indicial
exponents given by $\{0,1-\gamma\}$, $
\{0,1-\delta\}$, $\{0,1-\epsilon\}$ and $\{\alpha, \beta\}$,
respectively. 
The constants $a$, $\alpha$,
$\beta$,  $\gamma$ and $\delta$ are called singularity
parameters, whereas $q$ is
called accessory parameter
since it is not associated with the singular points or
their indicial exponents.

By defining the operator $M_{x}$ as
\begin{eqnarray}\label{operator}
M_{x}=x(x-1)(x-a)
\frac{\partial^{2}}{\partial x^2}+
\big[\gamma(x-1)(x-a)+\delta{x}(x-a)+
\epsilon{x}(x-1)\big]\frac{\partial}{\partial x}+\alpha \beta x
\end{eqnarray}
and, by interpreting this as an ordinary derivative operator, the
equation reads
\begin{eqnarray}\label{heun-2}
\big[M_{x}-q\big]H(x)=0,\qquad [a\neq 0\text{ or }1],
\end{eqnarray}
The invariance of the equation with respect to the replacement
of $\alpha$ by $\beta$ does not imply that
its solutions are symmetric in $\alpha$ and $\beta$;
it simply means that
the substitution of $\alpha$ for $\beta$
leads to another solution. The values $a=0$
and $a=1$ are excluded because in these cases
there are only three singular points and  then the equation
may be reduced to the Gauss hypergeometric equation
\begin{eqnarray}
\label{hypergeometric}
u(1-u)\frac{d^{2}F}{du^{2}}+\big[\mathrm{c}-(\mathrm{a}+\mathrm{b}+1)u
\big]\frac{dF}{du}-\mathrm{a}\mathrm{b}F=0,
\end{eqnarray}
where $u=0,1,\infty$ are regular
singular points with indicial exponents $\{0,1-c\}$, $\{0,\mathrm{c-a-b}\}$ and
${\mathrm\{a,b\}}$, respectively.

By keeping $\alpha$, $\gamma$ and $\delta$ fixed,
the confluence procedure is given by the limits
\begin{eqnarray}\label{confluence1}
a, \ \beta,\ q\to\infty\quad \mbox{such that}\quad \frac{\beta}{a} \to
\frac{\epsilon}{a}\to -\rho, \quad \frac{q}{a}\to -\sigma,
\end{eqnarray}
where $\rho$ and $\sigma$ are constants. This yields the
CHE \cite{arscott2}, or generalised spheroidal
wave equation \cite{leaver},
\begin{eqnarray}\label{confluence2}
x(x-1)\frac{d^2 H}{dx^2}+\big[-\gamma+(\gamma+\delta)x+\rho x(x-1)\big]
\frac{dH}{dx}+[\alpha\rho x-\sigma]H=0,
\end{eqnarray}
where $x=0$ and $x=1$ are regular singularities,
whereas $x=\infty$ is an irregular singularity.

On the other hand, $\mathcal{H}(x)$ is defined
by  \cite{lambe,erdelyi5,arscott2}
\begin{eqnarray}
\label{integral-heun}
\mathcal{H}(x)=\int_{y_{1}}^{y_{2}}w(x,y)\textsf{G}(x,y)H(y)dy
=\int_{y_{1}}^{y_{2}}y^{\gamma-1}
(1-y)^{\delta-1}\left( 1-\frac{y}{a}\right)^{\epsilon-1} \textsf{G}(x,y)H(y)dy,
\end{eqnarray}
where $H(x)$ represents a solution of
equation (\ref{heun}). Then, $\mathcal{H}(x)$
will be a solution of the Heun equation
if: (i) the kernel $\textsf{G}(x,y)$ is solution of
the partial differential equation
\begin{eqnarray}
\label{nucleus}
\left[M_{x}-{ M}_{y}\right] \textsf{G}(x,y)=0,
\end{eqnarray}
where $M_y$ is obtained by setting $x=y$ in the expression
for $M_x$, (ii) the integral (\ref{integral-heun})
exists and (iii) the limits of integration are so chosen that the
bilinear concomitant  $\textsf{P}(x,y)$, given by
\begin{eqnarray}
\label{concomitant-heun}
\textsf{P}(x,y)=
y^{\gamma}(1-y)^{\delta}\left(1-\frac{y}{a} \right) ^{\epsilon}\left[H(y)
\frac{\partial \textsf{G}(x,y)}{\partial y}-\textsf{G}(x,y)\frac{dH(y)}{d y}
\right],
\end{eqnarray}
fulfills the condition $\textsf{P}(x,y_1)=\textsf{P}(x,y_2)$.
In Appendix A we show how these equations
are obtained from the general
theory of integral relations \cite{ince1}.

By the choice given in Eq. (\ref{integral-heun})
for the weight function $w(x,y)$, equation (\ref{nucleus})
for the kernels is expressed in terms of the operator
$ M_x$ which appears in the Heun equation (\ref{heun})
and in terms of the functionally identical 
operator $ M_y$ obtained by setting $x=y$ in
$ M_x$. Then, in order to establish the transformations of the
kernels it is sufficient to demand that $ M_x$ and $ M_y$
transform in the same way. For the Heun equation these transformations 
will be inferred from the Maier transformations for the
Heun equation.

By using suitable weight functions
the above result holds also 
for the other equations of the Heun family,
that is, for confluent, double-confluent,
biconfluent and triconfluent Heun equations. Then, 
the transformation for the kernels 
may be inferred from the known transformations 
of each equation \cite{decarreaux2}. 
It seems that this connection has not been explored 
as yet \cite{ronveaux,kazakov1,slavyanov}.

However, the transformations 
become effective only if we know an initial kernel. For the Heun
equation, new kernels in terms of single hypergeometric
functions will be generated from the kernels
found by Lambe and Ward \cite{lambe}, while
kernels given by products of two hypergeometric
functions will arise from the ones found by Erd\'elyi \cite{erdelyi5}.
These afford initial kernels for the CHE by the 
limiting process (\ref{confluence1}). In addition to kernels
given by confluent hypergeometric functions, we
find kernels given by hypergeometric functions,
products of two confluent hypergeometric
functions, and products of one confluent hypergeometric
function and one hypergeometric function.

In section 2, firstly we present
the 8 so-called index or homotopic
transformations which do
not change the independent
variable $x$, and the 24 M\"obius or homographic transformations
which result from linear fractional substitutions of the
independent variable. Composition of such substitutions
gives the group of 192 transformations. After this, the 
transformations are extended to the
kernels of the equation, and these are used to generalise the 
kernels of Lambe-Ward and Erd\'elyi. 

In order to generate the full group 
by composition of homotopic and homographic transformations,
it is necessary to use the index transformations in Maier's form.
This remark is important for avoiding incorrect results. 
For example, the forms given in Refs. \cite{arscott2} and \cite{nist}
are inappropriate as we shall explain in 
section 2.1.

The kernels for the CHE are obtained in section 3, where we introduce 
as well the transformations of Eq. (\ref{confluence2}) and its kernels.
In section 4 we point out that even for the double-confluent Heun equation
(DCHE) it is possible to determine new kernels by using again a limiting process, 
and discuss how to transform certain solutions of the Heun equation 
into solutions useful for applications.
Appendix A provides a derivation of Eqs. (\ref{nucleus}) and
(\ref{concomitant-heun}), while 
Appendix B lists the M\"obius
transformations for kernels of the Heun equation.

\section*{2. Heun's equation}
First we examine the transformations
for the Heun equation, emphasising that
it is not allowed to permute the parameters
$\alpha_i$ and $\beta_i$ in the homotopic (index) transformations. Second, we
obtain a general prescription for the transformations
which preserve the equation for the kernels and write,
explicitly, the index transformations
for the kernels. In the third and fourth subsections, respectively,
we generalise the kernels given by hypergeometric
functions and by products of hypergeometric
functions.

\subsection*{2.1. Transformations of Heun's equation}

There are 24 (including the identity)
M\"obius substitutions of the independent variable
$x$ which leave the form of Heun's equation invariant,
in general after a change of the dependent variable.
They are given by fractional linear transformations
$x\mapsto$ $\varrho(x)=(Ax+B)/(Cx+D)$, $AD\neq BC$,
which map three of the points $0$, $1$, $a$ and
$\infty$ onto $0$, $1$, $\infty$. The expressions
for $\varrho(x)$ are displayed in the matrix 
\begin{eqnarray}\label{24fracional}
\left[
\begin{array}{cccccccccccc}
x\  &
\frac{x}{x-1}\ &
\frac{x}{x-a}\ &
1-x\ &
\frac{x-1}{x-a}\ &
\frac{a-x}{a};\ &
 \frac{1}{x}\ &
\frac{x-1}{x}\ &
%
\frac{x-a}{x}\ &
\frac{1}{1-x}\ &
\frac{x-a}{x-1}\ &
\frac{a}{a-x} \vspace{3mm}\\
%
%
\frac{x}{a}\ &
\frac{(a-1)x}{a(x-1)}\  &
\frac{(1-a)x}{x-a}\ &
\frac{1-x}{1-a}\ &
\frac{a(x-1)}{x-a}\ &
\frac{a-x}{a-1};\ &
\frac{a}{x}\ &
\frac{a(x-1)}{(a-1)x}\  &
\frac{x-a}{(1-a)x}\ &
\frac{1-a}{1-x}\ &
\frac{x-a}{a(x-1)}\ &
\frac{a-1}{a-x}
\end{array}\right]
\end{eqnarray}
where the elements in each column are proportional to one another
and, in each row, the elements after the semicolon
are the inverses of the elements before semicolon. 
For the identity, and for $(x-a)/(x-1)$, $a(x-1)/(x-a)$ and $a/x$ the other
singular point is mapped onto $a$, while for the remaining
cases it changes to \cite{arscott2,maier}
\begin{eqnarray}\label{a-i}
\frac{1}{a},\ \  1-a,  \ \ \frac{1}{1-a},
\ \ \frac{a}{a-1}, \ \ \frac{a-1}{a}.
\end{eqnarray}

Sometimes solutions for the
Heun equation are
denoted by $H(x)=Hl(a,q;\alpha,\beta,\gamma,\delta;x)$,
where $Hl$ means `Heun-local', that is, a solution
which converges in a region containing only one of the
four singular points \cite{arscott2,maier}.
For brevity, we drop the letter $l$,
writing $H(x)=H(a,q;\alpha,\beta,\gamma,\delta;x)$.
Then, the M\"obius substitutions permit onto map
a solution $H(x)$ into new solutions according to
\begin{eqnarray}\label{fraction}
H(a,q;\alpha,\beta,\gamma,\delta;x)\mapsto
f(x)H\left[\tilde{a},\tilde{q};\tilde{\alpha},\tilde{\beta},
\tilde{\gamma},\tilde{\delta};\varrho(x)\right],
\end{eqnarray}
where the prefactor $f(x)$ symbolises the transformation,
if any, of the dependent variable which brings the
differential equation with
the variable $\varrho(x)$ into
a Heun equation having parameters $\tilde{a}$, 
$\tilde{q}$, $\tilde{\alpha}$,
$\tilde{\beta}$,  $\tilde{\gamma}$ and $\tilde{\delta}$.
Depending on the transformation considered, we have
\begin{eqnarray}\label{prefactor-alfa}
 f(x)=1, \qquad x^{-\alpha},\qquad (1-x)^{-\alpha},
\qquad \left( 1-{x}/{a}\right)^{-\alpha},\text{ or }
\end{eqnarray}
\begin{eqnarray}\label{prefactor-beta}
 f(x)=1, \qquad x^{-\beta},\qquad (1-x)^{-\beta},
\qquad \left( 1-{x}/{a}\right)^{-\beta},\qquad
\end{eqnarray}
up to a multiplicative constant. The prefactor
$f(x)=1$ corresponds to the linear transformations, namely:
$\varrho(x)$ $=$ $1-x$, $(a-x)/a$, $x/a$, $(1-x)/(1-a)$ and $(a-x)/(a-1)$. 
The first form (\ref{prefactor-alfa}) is the one that will
be adopted in the present article.

On the other side, the index
transformations do not
change the independent variable.
They are given by 8 elementary power transformations of the
dependent variable \cite{arscott2,maier}, namely,
\begin{eqnarray}\label{elementary-2}
H(a,q;\alpha,\beta,\gamma,\delta;x)\mapsto
x^{\tau_{1}} (1-x)^{\tau_{2}}
\left( 1-{x}/{a}\right)^{\tau_{3}}
H(a,\tilde{q};\tilde{\alpha},\tilde{\beta},
\tilde{\gamma},\tilde{\delta};x),
\end{eqnarray}
where $\tau_{1}$, $\tau_{2}$ and  $\tau_{3}$
are the indicial exponents at $0$, $1$ and $a$, 
respectively, namely: $\tau_{1}=0$ or $1-\gamma$, $\tau_{2}=0$
or $1-\delta$, and $\tau_{3}=0$ or $1-\epsilon$.
Since there is no change of
the independent variable, the positions of the singular points
remain fixed,
in contrast with the fractional transformations.
For this reason, they are also called
homotopic transformations.

The composition of these two types of transformations
(elementary powers and fractional) generates
the group containing the 192 transformations
given in Maier's table \cite{maier}.
We refer to such transformations by
${M}_{i} \ (i=1,2,\cdots,192)$
following the order in which they
appear in the table, ${M}_{1}$ being the identity
transformation. Thence, by regarding the ${M}_{i}$ as operators,
the effects of both transformations on a
solution $H(x)$  are represented by
\begin{eqnarray}\label{fraction-2}
{M}_{i}H(x)=
{M}_{i}H(a,q;\alpha,\beta,\gamma,\delta;x)=
f_{i}(x)H\left[{a}_i,q_i;{\alpha}_i,{\beta}_i,
{\gamma}_i,{\delta}_i;\varrho_{i}(x)\right],
\end{eqnarray}
where $a_i$ is given by $a$ or one of the expressions
written in (\ref{a-i}), whereas $\varrho_{i}(x)$ is
one of the elements of the matrix (\ref{24fracional}).

In Maier's table the  elementary power transformations
appear in the entries $M_{1}-M_{4}$
and ${M}_{25}-M_{28}$,
but here they are denoted by $T_{i}$ ($i=1,\cdots,8$)
according to the correspondence
\begin{eqnarray}\label{arscott-maier}
\left[
\begin{matrix}
T_1\ &
T_2\ &
T_3\ &
T_4\ &
T_5\ &
T_6\ &
T_7\ &
T_8 &\vspace{3mm}\\
M_1\ &
M_{25}\ &
M_2\ &
M_{26}\ &
M_3\ &
M_{27}\ &
M_4 &
M_{28}
\end{matrix}
\right],
\end{eqnarray}
where $T_1$ is the identity: $T_1H(x)$$=$$H(a,q;\alpha,\beta,\gamma;\delta;x)$.
The transformations $T_2$, $T_3$ and $T_5$ are given by 
\begin{eqnarray}
\begin{array}{l}
T_{2}H(x)=
x^{1-\gamma}H\big[a,q-(\gamma-1)(\delta a+\epsilon);{\beta-\gamma+1},
\alpha-\gamma+1,2-\gamma,\delta;x \big],\vspace{2mm}\\
%
T_{3}H(x)= (1-x)^{1-\delta}H\big[a,q-(\delta-1)\gamma a;{\beta-\delta+1},
\alpha-\delta+1,\gamma,2-\delta;x \big],\vspace{2mm}\\
%
T_{5}H(x)=
\displaystyle\big[1-({x}/{a}) \big]^{1-\epsilon}
H\big[a,q-\gamma(\alpha+\beta-\gamma-\delta);
{-\alpha+\gamma+\delta},
-\beta+\gamma+\delta,
\gamma,\delta;x \big].
\end{array}
\end{eqnarray}
and are the generators
of the other $T_i$. 
In these transformations we cannot change the
order of the parameters $\alpha_i$ and $\beta_i$,
that is, we must read
\begin{eqnarray}\begin{array}{lll}
\text{for }T_2:\quad&\alpha_2=\beta-\gamma+1,&\quad \beta_2=\alpha-\gamma+1;\vspace{2mm}\\
\text{for }T_3:&\alpha_3=\beta-\delta+1,&\quad \beta_3=\alpha-\delta+1;\vspace{2mm}\\
\text{for }T_5:&\alpha_5=-\alpha+\gamma+\delta,&\quad \beta_5=-\beta+\gamma+\delta.
\end{array}
\end{eqnarray}
In effect, it is possible to obtain 
the 192 transformations only if $T_2$, $T_3$ and $T_5$ 
transform the prefactors
$x^{-\alpha}$, $(1-x)^{-\alpha}$ and $(1-x/a)^{-\alpha}$
of the M\"obius transformations
into $x^{-\beta}$, $(1-x)^{-\beta}$ and $(1-x/a)^{-\beta}$,
and {\it vice-versa}. For the other transformations, the
positions of $\alpha_i$ and $\beta_i$ result
from the compositions
\begin{eqnarray}
{T}_4H(x)=T_2T_3H(x),\ \
{T}_6H(x)={T}_2{T}_5H,\ \
{T}_7H(x)={T}_3{T}_5H(x),\ \
{T}_8H={T}_2{T}_3{T}_5H(x),\ \ 
\end{eqnarray}
where the order of the operators $T_i$ is irrelevant
on the right-hand side since $T_iT_j$$=$$T_jT_i$.
In spite of this, for the three transformations Sleeman and Kuznetsov \cite{nist} 
writes $\alpha_i$ and $\beta_i$ in the inverse order, 
while Arscott \cite{arscott2} inverts the order in $T_2$ and $T_3$.

The above order for $\alpha_i$ and $\beta_i$ 
is important regardless the composition among the two types of transformations.
This becomes apparent by considering an Erd\'elyi
solution in series of hypergeometric functions 
$F(\mathrm{a,b;c};x)$, given by \cite{erdelyi3}
\begin{eqnarray}\label{erdelyi-1944}
H_{1}(x)=\displaystyle\sum_{n=0}^{\infty}b_n^{(1)}\
F\left( n+\alpha,-n-\alpha-1+\gamma+\delta;
\gamma;x\right),
\end{eqnarray}
where the coefficients $b_n^{(1)}$ 
satisfy three-term recurrence relations. From this we can generate 
a subgroup constituted by 8 solutions by
writing $H_{i}(x)=T_iH_{1}(x)$. In particular, the solution with
$\alpha$ in the place of $\beta$ is 
\begin{eqnarray}\label{erd-2}
H_{3}(x)=T_3H_{1}(x)&=&\displaystyle(1-x)^{1-\delta}\sum_{n=0}^{\infty}b_n^{(3)}\
F\left( n+\beta+1-\delta,-n-\beta+\gamma;
\gamma;x\right)\nonumber\\&=&
\sum_{n=0}^{\infty}b_n^{(3)}\
F\left( n+\beta,-n-\beta-1+\gamma+\delta;
\gamma;x\right),
\end{eqnarray}
where the last equality follows from Eq. (\ref{euler-1})
written later on.
By interchanging $\alpha_3$ and $\beta_3$ in $T_3$, we would obtain 
the identity $H_{3}(x)=H_{1}(x)$ , that is, we would miss one
solution at least.

As aforementioned, we take the M\"obius
transformations with prefactors given in Eqs.
(\ref{prefactor-alfa}) as the basic ones. Then,
the substitutions (\ref{24fracional})
correspond to the following entries in Maier's table:
\begin{eqnarray}\label{24fracional-M}
\left[
\begin{array}{cccccccccccc}
{M}_1 &
{M}_5 &
{M}_{13}&
M_{49}&
M_{57} &
M_{101};&
 M_{145}&
 M_{53}&
 M_{97}&
 M_{149}&
 M_{105}&
 M_{157}\vspace{2mm}\\
M_9\ &
M_{21}\ &
M_{17}\ &
M_{61}\ &
M_{65}\ &
M_{109};\ &
M_{153}\ &
M_{69}\ &
M_{117}\ &
M_{161}\ &
M_{113}\ &
M_{165}
\end{array}\right],
\end{eqnarray}
where each column presents the same prefactor.

\subsection*{2.2. Transformations of kernel equation and notations}

Since Eq. (\ref{nucleus}) is independent of the
parameter $q$, a general kernel will be denoted by 
$\textsf{G}(x,y)$ $=$ $\textsf{G}(a;\alpha,\beta,\gamma,
\delta;x,y)$. Then, the transformations which
take the place of previous M\"obius and index
transformations are given, respectively, by the
mappings (symmetrical in $x$ and $y$)
\begin{eqnarray*}
&&\textsf{G}\left[ a;\alpha,\beta,\gamma,\delta;x,y\right]
\mapsto
f(x)f(y)\textsf{G}\left[\tilde{a};\tilde{\alpha},\tilde{\beta},
\tilde{\gamma},\tilde{\delta};\varrho(x),\varrho(y)\right],\\
%
&&\textsf{G}\left[ a;\alpha,\beta,\gamma,\delta;x,y\right] \mapsto
[xy]^{\tau_{1}} \left[ (1-x)(1-y)\right] ^{\tau_{2}}
\left[ \left( 1-\frac{x}{a}\right)
\left( 1-\frac{y}{a}\right)\right] ^{\tau_{3}}
\textsf{G}\left[ a;\tilde{\alpha},\tilde{\beta},
\tilde{\gamma},\tilde{\delta};x,y\right] ,
\end{eqnarray*}
where the prefactors $f(x)$ and $f(y)$, as well as
the fractional transformations
$\varrho(x)$ and $\varrho(y)$, are
formally the same which occur in the transformations
of the Heun equation. These substitutions
preserve the form of equation
$[M_x-M_y]\textsf{G}=0$ for the kernels
because all the parameters of the
operators $M_x$ and $M_y$
transform as in the Heun equation, and constant
terms corresponding to the transformations of $q$
cancel out. In terms of operators we rewrite
these transformations as
\begin{eqnarray}\label{fraction-G}
\mathcal{K}_{i}\textsf{G}(x,y)=
\mathcal{K}_{i}\textsf{G}(a;\alpha,\beta,\gamma,\delta;x,y)
=
f_{i}(x)f_{i}(y)\textsf{G}\big[a_i;{\alpha}_i,{\beta}_i,
{\gamma}_i,{\delta}_i;\varrho_{i}(x),
\varrho_{i}(y)\big],
\end{eqnarray}
where $\mathcal{K}_{i}$ is obtained from the
corresponding ${M}_{i}$ of Eq. (\ref{fraction-2}).

We split the operators $\mathcal{K}_{i}$ in two subgroups:
the operators denoted by  $N_i$ correspond to the 8 index
transformations $T_i$, and the operators denoted by ${K}_{i}$
correspond to the 24 M\"obius transformations.
Thus, the homotopic transformations  $N_i$ for the kernels are
\begin{eqnarray*}
N_1\textsf{G}(x,y)=\textsf{G}(x,y)=
\textsf{G}\left[a; {\alpha},\beta,\gamma,\delta;x,y\right]
\qquad[\text{Identity}],\hspace{3cm}
\end{eqnarray*}
\begin{eqnarray*}
N_2\textsf{G}(x,y)= &
(xy)^{1-\gamma}\textsf{G}\big[a;
{\beta-\gamma+1},\alpha-\gamma+1,2-\gamma,\delta;
x,y \big].\hspace{2.5cm}
\end{eqnarray*}
\begin{eqnarray*}
N_3\textsf{G}(x,y)&=& [(1-x)(1-y)]^{1-\delta}
\textsf{G}\big[a; {\beta-\delta+1},\alpha-\delta+1,\gamma,2-\delta;x,
 y\big],\hspace{.8cm}
\end{eqnarray*}
\begin{eqnarray*}
N_4\textsf{G}(x,y)&=& (xy)^{1-\gamma}[(1-x)(1-y]^{1-\delta}
\nonumber\\
%
&\times &
\textsf{G}\big[a;
{\alpha-\gamma-\delta+2},
\beta-\gamma-\delta+2,2-\gamma,2-\delta;x,y \big],\hspace{1.8cm}
\end{eqnarray*}
\begin{eqnarray*}
N_5\textsf{G}(x,y)=
\displaystyle\left[\left(1-\frac{x}{a} \right)
\left(1-\frac{y}{a} \right)\right]^{1-\epsilon}
\textsf{G}\big[a; {-\alpha+\gamma+\delta},
-\beta+\gamma+\delta,
\gamma,\delta;x, y\big],\ \ 
\end{eqnarray*}

\begin{eqnarray*}\displaystyle
N_6\textsf{G}(x,y)&=&\displaystyle (xy)^{1-\gamma}
\left[\left( 1-\frac{x}{a}\right)
\left( 1-\frac{y}{a}\right)\right]^{1-\epsilon}
\nonumber\\
%
&\times &\displaystyle
\textsf{G}\big[a;
{-\beta+\delta+1},
-\alpha+\delta+1,2-\gamma,\delta;x;y \big],\hspace{3.1cm}
\end{eqnarray*}
\begin{eqnarray*}
N_7\textsf{G}(x,y)&=&
\displaystyle [(1-x)(1-y)]^{1-\delta}
\left[\left(1-\frac{x}{a} \right)\left(1-\frac{y}{a} \right)\right]^{1-\epsilon}
\nonumber\\
%
&\times &
\textsf{G}\big[a;
{-\beta+\gamma+1},-\alpha+\gamma+1,
\gamma,2-\delta;x,y \big],\hspace{3.1cm}
\end{eqnarray*}
\begin{eqnarray*}
N_8\textsf{G}(x,y)&=&\displaystyle
 (xy)^{1-\gamma}[(1-x)(1-y)]^{1-\delta}
\left[\left(1-\frac{x}{a} \right)\left(1-\frac{y}{a}
\right)\right]^{1-\epsilon}\hspace{2cm}
\nonumber\\
%
&\times &\displaystyle
\textsf{G}\big[a; {2-\alpha},2-\beta,
2-\gamma,2-\delta;x,y \big].
\end{eqnarray*}
Notice that in fact $N_1$ is the identity only if
$\alpha\mapsto\alpha$ and $\beta\mapsto \beta$.
As in $T_i$, we cannot permute
$\alpha_i$ and $\beta_i$ in $N_2$,
$N_3$ and $N_5$  because
such transformations  must change
the exponents $\alpha$ of the prefactors (\ref{PREFACTORS})
(for fractional transformations) into $\beta$,
and \textit{vice-versa}.
For the other transformations, the
above positions of $\alpha_i$ and $\beta_i$ result
from the compositions
\begin{eqnarray}\label{N-compostas}
\begin{array}{ll}
N_4\textsf{G}(x,y)=N_2N_3\textsf{G}(x,y),&\qquad
N_6\textsf{G}(x,y)=N_2N_5\textsf{G}(x,y),\vspace{2mm}\\
%
N_7\textsf{G}(x,y)=N_3N_5\textsf{G}(x,y),&\qquad
N_8\textsf{G}(x,y)=N_2N_3N_5\textsf{G}(x,y),
\end{array}
\end{eqnarray}
where the order of the operators $N_i$ is irrelevant
on the right-hand side.
Thus, $N_2$, $N_3$ and $N_5$ are the generators
of the index transformations for the kernels.

On the other side, each of the 24
M\"obius transformations $M_i$
given in the matrix (\ref{24fracional-M})
is associated to a kernel transformation
denoted by $K_j$ ($j=1,\cdots,12$ for the
first row and $j=13,\cdots,24$ for the second) as 
\begin{eqnarray}\label{matriz-k}
\left[
\begin{array}{cccccccccccc}
K_1\ &
K_2\ &
K_3\ &
K_4\ &

K_5\  &
K_6;\ &
K_7\ &
K_8 \ &
K_9 \ &
K_{10}\ &
K_{11}\  &
K_{12}\vspace{3mm}\\
K_{13}\  &
K_{14}\  &
K_{15}\ &
K_{16}\ &
K_{17}\ &
K_{18};\ &
K_{19}\ &
K_{20} \ &
K_{21} \ &
K_{22}\ &
K_{23}\ &
K_{24}\
\end{array}\right].
\end{eqnarray}
As in the homotopic transformations, first we must  find
the transformation $M_i$ for $H(x)$
in Maier's table and, then, write the kernel
transformation $K_i$ by using
the prescription given in Eq. (\ref{fraction-G}).
The 24 expressions for $K_i$ are written down in Appendix B and
will be used in following subsections. The prefactors
for those transformations are
\begin{eqnarray}\label{PREFACTORS}
 1, \qquad (zy)^{-\alpha},\qquad \left[(1-x)(1-y)\right]^{-\alpha},
\qquad \left[\left( 1-\frac{x}{a}\right)\left( 1-\frac{y}{a}\right)\right]^{-\alpha}.
\end{eqnarray}
Transformations having prefactors with
exponent $\beta$ are generated
from the ones of Appendix B by applying $N_2$
when the prefactor is $(zy)^{-\alpha}$, $N_3$ when $\left[(1-x)(1-y)\right]^{-\alpha}$
and $N_5$ when
$\left[\left( 1-{x}/{a}\right)\left( 1-{y}/{a}\right)\right]^{-\alpha}$.

We could denote an initial kernel
by $\textsf{G}_{1}^{1}(x,y)$,
and by $\textsf{G}_{i}^{j}(x,y)$ the kernels
obtained from  $\textsf{G}_{1}^{1}(x,y)$
by using $N_{i}$ and ${K}_{j}$, where
\begin{eqnarray*}
\text{in }\textsf{G}_{i}^{j}(x,y): \left\lbrace  \begin{array}{l}
j \text{ indicates M\"obius transformation, } {K}_{j},\vspace{2mm}\\
i \text{ indicates index transformation, } 
N_{i},                                          
\end{array} \right.
\end{eqnarray*}
being necessary to specify the transformation
applied in the first place since in general
index and M\"obius transformations do not
commute. However, this notation
is not sufficient because we will consider a set
of initial kernels rather than a single kernel. For
the Lambe-Ward case, we start with six kernels given
by distinct hypergeometric functions and use
the notation $\textsf{G}_{1}^{1(k)}$ where
$k$ runs from 1 to 6; for the Erd\'elyi case we take 36
products of hypergeometric functions and then
the initial set is denoted by $\textsf{G}_{1}^{1(k,l)}$.
Thence, the actual notation will be
\begin{eqnarray}
\textsf{G}_{i}^{j(k)}(x,y)\text{ for Lambe-Ward-type kernels };\quad
\textsf{G}_{i}^{j(k,l)}(x,y)\text{  for Erd\'elyi-type},
\end{eqnarray}
where the indices inside parentheses are not affected by
the application of the transformations $N_i$ and $K_{j}$.

The Lambe-Ward kernels $\textsf{G}_{1}^{1(k)}(x,y)$ defined
in Eq. (\ref{def-Lambe}) and
Erd\'elyi kernels $\textsf{G}_{1}^{1(k,l)}(x,y)$ defined
in Eq. (\ref{erdelyi-2}) 
are the initial kernels which are 
obtained by solving directly the kernel equation.
We will find that $K_j$ with $j=2,\cdots,6$ are the only
effective M\"obius transformations. These lead to
five additional sets of Lambe-Ward-type kernels, denoted and
obtained as
\begin{eqnarray}\label{lambe}
\begin{array}{lll}
\textsf{G}_{1}^{2(k)}=K_2\textsf{G}_{1}^{1(k)},&\qquad
\textsf{G}_{1}^{3(k)}=K_3\textsf{G}_{1}^{1(k)},&\qquad
\textsf{G}_{1}^{4(k)}=K_4\textsf{G}_{1}^{1(k)},\vspace{2mm}\\
\textsf{G}_{1}^{5(k)}=K_5\textsf{G}_{1}^{1(k)},&\qquad
\textsf{G}_{1}^{6(k)}=K_6\textsf{G}_{1}^{1(k)},&\qquad[k=1,2,\cdots,6].
\end{array}
\end{eqnarray}
Similarly, the new Erd\'elyi-type kernels are
\begin{eqnarray}\label{erdely}
\begin{array}{lll}
\textsf{G}_{1}^{2(k,l)}=K_2\textsf{G}_{1}^{1(k,l)},&\qquad
\textsf{G}_{1}^{3(k,l)}=K_3\textsf{G}_{1}^{1(k,l)},&\qquad
\textsf{G}_{1}^{4(k,l)}=K_4\textsf{G}_{1}^{1(k,l)},\vspace{2mm}\\
\textsf{G}_{1}^{5(k,l)}=K_5\textsf{G}_{1}^{1(k,l)},&\qquad
\textsf{G}_{1}^{6(k,l)}=K_6\textsf{G}_{1}^{1(k,l)},&\qquad[k,l=1,2,\cdots,6].
\end{array}
\end{eqnarray}
In each case, the subscript could assume eight values
when we apply the homotopic transformations $N_i$.
Nevertheless, we will find that only three of the $N_i$ 
are effective due to fact
that one of the generators $N_2$, $N_3$, $N_5$
becomes equivalent to the identity or two of them
are equivalent to each other. 
Thus, there are only four values for the subscripts.

\subsection*{2.3. Generalisation of Lambe-Ward's kernels}

The Lambe-Ward as well as the Erd\'elyi kernels
are given by  hypergeometric functions
$F(\mathrm{a,b;c};u)=F(\mathrm{b,a;c};u)$.
In fact, in the vicinity of the singular points $0$,
$1$ and $\infty$, the formal solutions for the hypergeometric
equation (\ref{hypergeometric}) are \cite{erdelii}, respectively,
\letra
\begin{eqnarray}\label{hiper-1}
\begin{array}{l}
F^{(1)}(u)=F\left( \mathrm{a,b;c;}
u\right) ,
\qquad
F^{(2)}(u)=u^{1-\mathrm{c}}
F\left( \mathrm{a+1-c,b+1-c;
2-c};u\right);\ \ 
\end{array}
\end{eqnarray}
\begin{eqnarray}\label{hiper-2}
\begin{array}{l}
F^{(3)}(u)=F\left( \mathrm{a,b;
a+b+1-c};1-u \right),\vspace{2mm}\\
%
F^{(4)}(u)=\left( 1-u\right)^{\mathrm{c-a-b}}
F\left(\mathrm{c- a,c-b;
1+c-a-b};1-u\right);
\end{array}
\hspace{2.5cm}
\end{eqnarray}
\begin{eqnarray}\label{hiper-3}
\begin{array}{l}
F^{(5)}(u)=u^{-\mathrm{a}}
F\left( \mathrm{a,a+1-c;
a+1-b};1/u\right),\vspace{2mm}\\
F^{(6)}(u)=u^{-\mathrm{b}}
F\left(\mathrm{b+1-c,b;
b+1-a};{1}/{u}\right).
\end{array}
\hspace{4.75cm}
\end{eqnarray}
Each of these functions may be written in four forms
by using the relations
\antiletra
\begin{eqnarray}\label{euler-1}
F(\mathrm{a,b;c;u})=(1-u)^{\mathrm{c-a-b}}
F(\mathrm{c-a,c-b;c;u}),\ \
\end{eqnarray}
\begin{eqnarray}\label{euler-2}
F(\mathrm{a,b;c;u})=(1-u)^{\mathrm{-a}}
F\left[ \mathrm{a,c-b;c;{u}/({u-1}})\right] .
\end{eqnarray}

The Lambe and Ward
kernels \cite{lambe} in terms of single
hypergeometric functions are obtained by taking $u=xy/a$
and $\textsf{G}(x,y)=F(u)$

in $\left[M_{x}-{ M}_{y}\right] \textsf{G}(x,y)=0$. Hence,
$F(u)$ satisfies equation
(\ref{hypergeometric}) with $\mathrm{a}=\alpha$,
$\mathrm{b}=\beta$ and $\mathrm{c}=\gamma$. Thus,
the six formal kernels have the form
\begin{eqnarray}\label{def-Lambe}
\textsf{G}_{1}^{1(i)}(x,y)=F^{(i)}\left(\frac{xy}{a}\right), \quad
[\mathrm{a}=\alpha,\ \mathrm{b}=\beta,\ \mathrm{c}=\gamma]
\end{eqnarray}
In this manner, up to a multiplicative constant, the initial
set of kernels is given by
\letra
\begin{eqnarray}\begin{array}{l}\displaystyle
\textsf{G}_{1}^{1(1)}(x,y)=
F\left( \alpha,\beta;\gamma;
\frac{xy}{a}\right) ,
\vspace{2mm}\\
\displaystyle
\textsf{G}_{1}^{1(2)}(x,y)=
(xy)^{1-\gamma}
F\left( \alpha+1-\gamma,\beta+1-\gamma;
2-\gamma;\frac{xy}{a}\right);
\end{array}\hspace{2cm}
\end{eqnarray}
\begin{eqnarray}\begin{array}{l}\displaystyle
\textsf{G}_{1}^{1(3)}(x,y)=
F\left( \alpha,\beta;
\alpha+\beta+1-\gamma;1-\frac{xy}{a} \right),\vspace{2mm}\\
\displaystyle
\textsf{G}_{1}^{1(4)}(x,y)=
\left( 1-\frac{xy}{a}\right)^{\gamma-\alpha-\beta}
F\left(\gamma- \alpha,\gamma-\beta;
1+\gamma-\alpha-\beta;1-\frac{xy}{a}\right);
\end{array}
\end{eqnarray}
\begin{eqnarray}\begin{array}{l}\displaystyle
\textsf{G}_{1}^{1(5)}(x,y)=
(xy)^{-\alpha}
F\left( \alpha,\alpha+1-\gamma;
\alpha+1-\beta;\frac{a}{xy}\right),\vspace{2mm}\\
\displaystyle
\textsf{G}_{1}^{1(6)}(x,y)=
(xy)^{-\beta}
F\left(\beta+1-\gamma,\beta;
\beta+1-\alpha;\frac{a}{xy}\right).
\end{array}\hspace{2.55cm}
\end{eqnarray}

By using this set of initial kernels, some of the kernel
transformations become superfluous.
In effect, by the transformations $N_i$
we could obtain a subgroup containing eight
sets. However, for the present
case $N_2$ is ineffective since
\begin{eqnarray*}
N_2\left(
\textsf{G}_1^{1(1)},
\textsf{G}_1^{1(2)},
\textsf{G}_1^{1(3)},
\textsf{G}_1^{1(4)},
\textsf{G}_1^{1(5)},
\textsf{G}_1^{1(6)}\right)=
\left(
\textsf{G}_1^{1(2)},
\textsf{G}_1^{1(1)},
\textsf{G}_1^{1(3)},
\textsf{G}_1^{1(4)},
\textsf{G}_1^{1(6)},
\textsf{G}_1^{1(5)}\right),
\end{eqnarray*}
that is, $N_2$ simply rearranges in a different order
the previous kernels.  In this sense the generator $N_2$ is equivalent
to the identity $N_1$ and, so, the index transformations
can generate only three additional sets due to composition
relations (\ref{N-compostas}). In fact we find that
\begin{eqnarray*}
N_3\textsf{G}_1^{1(i)}\Leftrightarrow N_4\textsf{G}_1^{1(i)},\qquad
N_5\textsf{G}_1^{1(i)}\Leftrightarrow N_6\textsf{G}_1^{1(i)},\qquad
N_7\textsf{G}_1^{1(i)}\Leftrightarrow N_8\textsf{G}_1^{1(i)}.
\end{eqnarray*}
Therefore, it is sufficient to use the
transformations $N_3$, $N_5$ and
$N_7$ to produce three additional sets, namely:
$\textsf{G}_{3}^{1(k)}$, $\textsf{G}_{5}^{1(k)}$
and $\textsf{G}_{7}^{1(k)}$.
The eight kernels $\textsf{G}_i^{1(1)}$ and
$\textsf{G}_i^{1(2)}$ ($i=1,3,5,7$) coincide with the ones
given by Lambe and Ward.
Now, we regard the generalisations arising from the M\"obius
transformations  $K_j$. By one side we find
\begin{eqnarray*}
{K}_{13}\left(
\textsf{G}_1^{1(1)},
\textsf{G}_1^{1(2)},
\textsf{G}_1^{1(3)},
\textsf{G}_1^{1(4)},
\textsf{G}_1^{1(5)},
\textsf{G}_1^{1(6)}\right)=
\left(
\textsf{G}_1^{1(1)},
\textsf{G}_1^{1(2)},
\textsf{G}_1^{1(3)},
\textsf{G}_1^{1(4)},
\textsf{G}_1^{1(5)},
\textsf{G}_1^{1(6)}\right),
\end{eqnarray*}
that is, the transformations ${K}_{1}$ (identity)
and ${K}_{13}$ of
the first column of (\ref{matriz-k})
are equivalent, a fact that holds
for the transformations of any column.
Thus we have to take into account
only the twelve transformations of the first row.
However, we find as well that
\begin{eqnarray*}
K_{7}\left(
\textsf{G}_1^{1(1)},
\textsf{G}_1^{1(2)},
\textsf{G}_1^{1(3)},
\textsf{G}_1^{1(4)},
\textsf{G}_1^{1(5)},
\textsf{G}_1^{1(6)}\right)=
\left(
\textsf{G}_1^{1(5)},
\textsf{G}_1^{1(6)},
\textsf{G}_1^{1(3)},
\textsf{G}_1^{1(4)},
\textsf{G}_1^{1(1)},
\textsf{G}_1^{1(2)}\right)
\end{eqnarray*}
that is, $K_{7}$ rearranges the initial kernels.
This could be expected because $K_{1}\mapsto$
$K_{7}$ corresponds to the
inversions $(x,y)\mapsto(1/x,1/y)$ which are already
incorporated into the original set. This is the same
for the other transformations
corresponding to inverted mappings,
that is, $K_i\Leftrightarrow K_{i+6}$ ($i=1,\cdots,6$).
Therefore, we may apply only the transformations
$K_{2}$, $K_{3}$, $K_{4}$, $K_{5}$ and
$K_{6}$ on $\textsf{G}_1^{1(k)}$ according to Eq. (\ref{lambe}) in order to
find new sets, each one containing
six hypergeometric functions.
To determine which are the
transformations $N_i$ that are
suitable to generate new kernels, it is sufficient to regard the pair
$\left( \textsf{G}_1^{j(1)},\textsf{G}_1^{j(2)}\right) $;
the other pairs may be obtained by replacing
the $F(\mathrm{a,b;c;}u)$
which appears in $\textsf{G}_1^{j(1)}$ by the
hypergeometric functions written in Eqs. (\ref{hiper-2}-c).

Thus, from $K_2$ we find
\antiletra
\begin{eqnarray}\begin{array}{l}\displaystyle
\textsf{G}_{1}^{2(1)}(x,y)=
\left[ (1-x)(1-y)\right] ^{-\alpha}F\left[ \alpha,1+\alpha-\delta;\gamma;
\frac{(a-1)xy}{a(x-1)(y-1)}\right] ,\vspace{3mm}\\
\displaystyle
\textsf{G}_{1}^{2(2)}(x,y)=
\left[ (1-x)(1-y)\right]^{\gamma-\alpha-1} (xy)^{1-\gamma}\times
\vspace{3mm}\\
\hspace{2.8cm}
F\left[ \alpha+1-\gamma,2+\alpha-\gamma-\delta;2-\gamma;
\frac{(a-1)xy}{a(x-1)(y-1)}\right],
\end{array}\end{eqnarray}
together with the two pairs generated by using
the hypergeometric functions (\ref{hiper-2}-c), 
as explained above. For this case, the generators $N_2$ and $N_3$
are equivalent to each other and, consequently, the $N_i$
afford only three additional sets.  We find
\begin{eqnarray*}
N_1\textsf{G}_1^{2(i)}\Leftrightarrow N_4\textsf{G}_1^{2(i)},
\quad
N_2\textsf{G}_1^{2(i)}\Leftrightarrow N_3\textsf{G}_1^{2(i)},
\quad
N_5\textsf{G}_1^{2(i)}\Leftrightarrow N_8\textsf{G}_2^{1(i)},
\quad
N_6\textsf{G}_1^{1(i)}\Leftrightarrow N_7\textsf{G}_1^{1(i)}.
\end{eqnarray*}
Thence we may use only the transformations $N_2$, $N_5$
and $N_6$. On the other hand,
by using $K_{3}$, we get
\begin{eqnarray}\begin{array}{l}
\displaystyle
\textsf{G}_{1}^{3(1)}(x,y)=
\left[ \left(1-\frac{x}{a}\right)
\left(1-\frac{y}{a}\right)\right]^{-\alpha}
F\left[ \alpha,\gamma+\delta-\beta;
\gamma;\frac{(1-a)xy}{(a-x)(a-y)}\right] ,\vspace{3mm}\\
\displaystyle
\textsf{G}_{1}^{3(2)}(x,y)=[xy]^{1-\gamma}
\left[ \left(1-\frac{x}{a}\right)\left(1-\frac{y}
{a}\right)\right]^{\gamma-1-\alpha}
\times\vspace{3mm}\\
\displaystyle\hspace{3.3cm}
F\left[ \alpha+1-\gamma,\delta+1-\beta;
2-\gamma;\frac{(1-a)xy}{(a-x)(a-y)}\right].
\end{array}
\end{eqnarray}
This time $N_1\Leftrightarrow N_6$,
$N_2\Leftrightarrow N_5$, $N_3\Leftrightarrow N_8$
and $N_4\Leftrightarrow N_7$. Thus,
the transformations $N_2$, $N_3$
and $N_4$ are sufficient. The transformation $K_{4}$ yields
\begin{eqnarray}\begin{array}{l}\displaystyle
\textsf{G}_{1}^{4(1)}(x,y)=
F\left[ \alpha,\beta;\delta;\frac{(x-1)(y-1)}{1-a}\right] ,
\vspace{3mm}\\
\displaystyle
\textsf{G}_{1}^{4(2)}(x,y)=
\left[ (1-x)(1-y)\right]^{1-\delta}
F\left[ \alpha+1-\delta,\beta+1-\delta;2-\delta;\frac{(x-1)(y-1)}{1-a}\right].
\end{array}
\end{eqnarray}
Since $N_1\Leftrightarrow N_3$,
$N_2\Leftrightarrow N_4$, $N_5\Leftrightarrow N_7$
and $N_6\Leftrightarrow N_8$,
we can choose only $N_2$, $N_5$ and $N_6$. By $K_{6}$ we get
\begin{eqnarray}
\label{lambe-6}
\begin{array}{l}
\displaystyle
\textsf{G}_{1}^{5(1)}(x,y)=
\left[\left( 1-\frac{x}{a}\right)\left( 1-\frac{y}{a}\right)
\right] ^{-\alpha}
F\left[ \alpha,\gamma+\delta-\beta;\delta;
\frac{a(x-1)(y-1)}{(x-a)(y-a)}\right] ,\vspace{3mm}\\
\displaystyle
\textsf{G}_{1}^{5(2)}(x,y)=
\left[(1-x)(1-y) \right]^{1-\delta}
\left[\left( 1-\frac{x}{a}\right)\left( 1-\frac{y}{a}\right)
\right] ^{\delta-1-\alpha}\times\vspace{3mm}\\
\displaystyle
\hspace{3.3cm}
F\left[ \alpha+1-\delta,\gamma+1-\beta;2-\delta;
\frac{a(x-1)(y-1)}{(x-a)(y-a)}\right],
\end{array}
\end{eqnarray}
together with the sets obtained by applying $N_2$, $N_3$
and $N_4$, because $N_1\Leftrightarrow N_7$,
$N_2\Leftrightarrow N_8$, $N_3\Leftrightarrow N_5$
and $N_4\Leftrightarrow N_6$ for this subgroup. Finally,
the transformation $K_{6}$ leads to
\begin{eqnarray}\label{lambe-8}
\begin{array}{l}
\displaystyle
\textsf{G}_{1}^{6(1)}(x,y)=
F\left[ \alpha,\beta;\epsilon;
\frac{(x-a)(y-a)}{a(a-1)} \right] ,
\vspace{3mm}\\
\displaystyle
\textsf{G}_{1}^{6(2)}(x,y)=
\left[\left(1-\frac{x}{a} \right) \left( 1-\frac{y}{a}\right)
\right]^{1-\epsilon}
F\left[ \gamma+\delta-\alpha,\gamma+\delta-\beta;2-\epsilon;
\frac{(x-a)(y-a)}{a(a-1)} \right],
\end{array}
\end{eqnarray}
and the sets resulting from this by  $N_2$, $N_3$
and $N_4$ since now $N_1\Leftrightarrow N_5$,
$N_2\Leftrightarrow N_6$, $N_3\Leftrightarrow N_7$
and $N_4\Leftrightarrow N_8$.

In summary, the substitutions of the independent
variables  that convert
the kernel equation (\ref{nucleus})
into a hypergeometric equation
are $(x,y)\mapsto u(x,y)$, where $ u(x,y)$ is given by
\begin{equation}
\frac{xy}{a}, \ 
\frac{(a-1)xy}{a(x-1)(y-1)}, \ 
\frac{(1-a)xy}{(a-x)(a-y)}, \ 
%
\frac{(x-1)(y-1)}{1-a},\ 
\frac{a(x-1)(y-1)}{(x-a)(y-a)},\ 
\frac{(x-a)(y-a)}{a(a-1)}.
\end{equation}
In general it is necessary to
perform a substitution of the dependent
variable as well.

\subsection*{2.4. Generalisation of Erd\'elyi's kernels}
These are given by products of hypergeometric functions
containing an arbitrary separation constant $\lambda$.
When this constant is appropriately chosen, we recover
the Lambe-Ward-type kernels.
The Erd\'elyi kernels are obtained by rewriting
equation (\ref{nucleus})
in terms of the independent variables \cite{erdelyi5}
\begin{eqnarray}\label{erdelyi-1}
\xi=\frac{xy}{a},\qquad \zeta=\frac{(x-a)(y-a)}{(1-a)(xy-a)},
\end{eqnarray}
and, then, by accomplishing separation of variables
in the resulting equation. It turns out that
\begin{eqnarray}\label{erdelyi}
\textsf{G}(x,y)=\big(1-\xi\big)^{-\lambda}P(\xi)
\ Q(\zeta),
\end{eqnarray}
where $P(\xi)$ and $Q(\zeta)$ satisfy the hypergeometric
equation (\ref{hypergeometric}) with the following sets of
parameters:
\begin{eqnarray}\label{PQ}
\begin{array}{llll}
 P(\xi):&
\mathrm{a}=\alpha-\lambda,\quad&
\mathrm{b}=\beta-\lambda,&
\mathrm{c}=\gamma;\vspace{3mm}\\
Q(\zeta):&
{\mathrm{a}}=\lambda,&
{\mathrm{b}}=\alpha+\beta-\gamma-\lambda,\quad&
{\mathrm{c}}=\epsilon.
    \end{array}
\end{eqnarray}

To show this, firstly we perform
the substitutions (\ref{erdelyi-1}) in Eq. (\ref{hypergeometric}). We get the equation
\begin{eqnarray*}
&&(1-\xi)\left\{\xi(1-\xi)\frac{\partial^2\textsf{G}}{\partial \xi^2}
+\Big[\gamma-(\alpha+\beta+1)\xi
 \Big] \frac{\partial\textsf{G}}{\partial \xi}
-\alpha\beta\textsf{G}\right\}+\\
&&\zeta(1-\zeta)\frac{\partial^2\textsf{G}}{\partial \zeta^2}
+\Big[\epsilon-(\delta+\epsilon)\zeta
 \Big] \frac{\partial\textsf{G}}{\partial \zeta}=0,
\end{eqnarray*}
which, by the separation of variables $\textsf{G}(\xi,\zeta)$
$=$ $\bar{P}(\xi)Q(\zeta)$, becomes
\begin{eqnarray*}
&&\frac{(1-\xi)}{\bar{P}}
\left\{\xi(1-\xi)\frac{d^2\bar{P}}{d \xi^2}
+\Big[\gamma-(\gamma+\delta+\epsilon)\xi
 \Big] \frac{d\bar{P}}{d \xi}
-\alpha\beta\bar{P}\right\}+\\
&&\frac{1}{Q}
\left\lbrace \zeta(1-\zeta)\frac{d^2Q}{d \zeta^2}
+\Big[\epsilon-(\delta+\epsilon)\zeta
 \Big] \frac{d Q}{d \zeta}\right\rbrace =0.
\end{eqnarray*}
Denoting the separation constant
by $\lambda(\alpha+\beta-\gamma-\lambda)$, we obtain
%
\begin{eqnarray}
&& \zeta(1-\zeta)\frac{d^2Q}{d \zeta^2}
+\Big[\epsilon-(\delta+\epsilon)\zeta
 \Big] \frac{d Q}{d \zeta}-
\lambda(\alpha+\beta-\gamma-\lambda)Q =0,\\
%
&&\xi(1-\xi)\frac{d^2\bar{P}}{d \xi^2}
+\Big[\gamma-(\gamma+\beta+1)\xi
 \Big] \frac{d\bar{P}}{d \xi}-
\left[ \alpha\beta-
\frac{\lambda(\alpha+\beta-\gamma-\lambda)}{1-\xi}
\right] \bar{P}=0.\nonumber
\end{eqnarray}
The additional substitution $\bar{P}=(1-\xi)^{-\lambda} P$
leads to
\begin{eqnarray}
\xi(1-\xi)\frac{d^2{P}}{d \xi^2}
+\Big[\gamma-(\alpha+\beta+1-2\lambda)\xi
 \Big] \frac{d{P}}{d \xi}-
(\alpha-\lambda)(\beta-\lambda){P}=0.
\end{eqnarray}
In this manner, $P(\xi)$ and $Q(\zeta)$ satisfy
hypergeometric equations with the parameters
given in (\ref{PQ}).
If $\lambda=0$, we can take $Q(\zeta)$ $=$ constant
in order to recover the Lambe-Ward kernels,
$\textsf{G}_{1}^{1(i)}$. On the
other hand, for  $\lambda=\alpha$ and $P(\xi)$ $=$
constant, the resulting kernels belong to
the generalised Lambe-Ward
kernels $\textsf{G}_{i}^{6(j)}(x,y)$ which accompany
the kernels (\ref{lambe-6}), since by
relation (\ref{euler-2}) the arguments
of the hypergeometric functions take the form
$u= {(x-a)(y-a)}/[{a(x-1)(y-1)]}$, $1-u$ or $1/u$.

For $P(\xi)$ we select two hypergeometric
functions in the neighbourhood
of each singular point, as in the
Lambe-Ward case. Since for $Q(\zeta)$
there are six possibilities as well, we write the initial
set as
\begin{eqnarray}\label{erdelyi-2}
\textsf{G}_{1}^{1(k,l)}(x,y)=\big(1-\xi\big)^{-\lambda}P^{(k)}(\xi)
\ Q^{(l)}(\zeta),
\end{eqnarray}
where $P^{(k)}(\xi)$ and $Q^{(l)}(\zeta)$  are obtained from
the six hypergeometric functions (\ref{hiper-1}-c), having parameters
specified in (\ref{PQ}). Explicitly, we find

\letra
\begin{eqnarray}\label{erdelyi-a}
\begin{array}{l}
\textsf{G}_{1}^{1(1,l)}(x,y)=
\displaystyle
\left(1-\frac{xy}{a}\right) ^{-\lambda}
F\left[ \alpha-\lambda,
\beta-\lambda;\gamma;\frac{xy}{a}\right]
Q^{(l)}(\zeta),\vspace{3mm}\\
\textsf{G}_{1}^{1(2,l)}(x,y)=
\displaystyle
\left( xy\right)^{1-\gamma} \left(1-\frac{xy}{a}\right) ^{-\lambda}
F\left[\alpha+1-\gamma-\lambda,
\beta+1-\gamma-\lambda;2-\gamma;\frac{xy}{a}\right]
Q^{(l)}(\zeta);
\end{array}
\end{eqnarray}
\begin{eqnarray}\label{erdelyi-b}
\begin{array}{l}
\displaystyle
\textsf{G}_{1}^{1(3,l)}(x,y)=
\left( 1-\frac{xy}{a}\right) ^{-\lambda}
F\left[ \alpha-\lambda,
\beta-\lambda;1+\alpha+\beta-\gamma-2\lambda;
1-\frac{xy}{a}\right]
Q^{(l)}(\zeta),\vspace{3mm}\\
\textsf{G}_{1}^{1(4,l)}(x,y)=
\displaystyle
\left( 1-\frac{xy}{a}\right)^{\gamma-\alpha-\beta+\lambda}
\times\vspace{3mm}\\
\hspace{2cm}\displaystyle
F\left[ \gamma-\alpha+\lambda,
 \gamma-\beta+\lambda;
1+\gamma-\alpha-\beta+2\lambda;
1-\frac{xy}{a}\right]
Q^{(l)}(\zeta);
\end{array}\hspace{.95cm}
\end{eqnarray}
\begin{eqnarray}\label{erdelyi-c}
\begin{array}{l}
\textsf{G}_{1}^{1(5,l)}(x,y)=
\displaystyle
\left(xy \right)^{\lambda-\alpha}
\left( 1-\frac{xy}{a}\right)^{-\lambda}
F\left[\alpha-\lambda,
\alpha+1-\gamma-\lambda;
1+\alpha-\beta;\dfrac{a}{xy}\right]Q^{(l)}(\zeta)
,\vspace{3mm}\\
\displaystyle
\textsf{G}_{1}^{1(6,l)}(x,y)=
\displaystyle
\left(xy \right)^{\lambda-\beta}
\left( 1-\frac{xy}{a}\right)^{-\lambda}
F\left[\beta-\lambda,
\beta+1-\gamma-\lambda;
1+\beta-\alpha;\dfrac{a}{xy}\right]
Q^{(l)}(\zeta),
\end{array}\hspace{.45cm}
\end{eqnarray}
where
\begin{eqnarray*}
Q^{(1)}(\zeta)=F\left[\lambda,
\alpha+\beta-\gamma-\lambda;\epsilon;
 \frac{(x-a)(y-a)}{(1-a)(xy-a)}\right],\hspace{5.4cm}
\end{eqnarray*}
\begin{eqnarray*}
Q^{(2)}(\zeta)=\left[\frac{(x-a)(y-a)}{(1-a)(xy-a)} \right]^{1-\epsilon}
F\left[\lambda+1-\epsilon,
\delta-\lambda;2-\epsilon;
 \frac{(x-a)(y-a)}{(1-a)(xy-a)}\right], \hspace{1.6cm}
\end{eqnarray*}
\begin{eqnarray*}\label{Q3}
Q^{(3)}(\zeta)=F\left[\lambda,
\alpha+\beta-\gamma-\lambda;\delta;
 1-\frac{(x-a)(y-a)}{(1-a)(xy-a)}\right],\hspace{4.7cm}
\end{eqnarray*}
\begin{eqnarray*}
Q^{(4)}(\zeta)=
\left[1-\frac{(x-a)(y-a)}{(1-a)(xy-a)} \right]^{1-\delta} F\left[\epsilon-\lambda,
1+\lambda-\delta;2-\delta;
 1-\frac{(x-a)(y-a)}{(1-a)(xy-a)}\right],\quad
\end{eqnarray*}
\begin{eqnarray*}
Q^{(5)}(\zeta)=
\left[\frac{(x-a)(y-a)}{(1-a)(xy-a)} \right]^{-\lambda} F\left[\lambda,
1+\lambda-\epsilon;1+2\lambda-\alpha-\beta;
 \frac{(1-a)(xy-a)}{(x-a)(y-a)}\right],\hspace{.8cm}
\end{eqnarray*}
\begin{eqnarray}
Q^{(6)}(\zeta)&=&
\left[\frac{(x-a)(y-a)}{(1-a)(xy-a)} \right]^{\lambda+\gamma-\alpha-\delta}
\nonumber\\
%
&\times&F\left[\delta-\lambda,
\alpha+\beta-\gamma+\lambda;1-2\lambda+\alpha+\beta;
 \frac{(1-a)(xy-a)}{(x-a)(y-a)}\right].\hspace{2.8cm}
\end{eqnarray}

We can generate additional
kernels with the same arguments for the
hypergeometric functions by applying
the index transformations $N_i$.
However, as in the case of the Lambe-Ward kernels, we find that
\begin{eqnarray*}
N_1G_1^{1(k,l)}\Leftrightarrow N_2G_1^{1(k,l)},\ 
N_3G_1^{1(k,l)}\Leftrightarrow N_4G_1^{1(k,l)},\ 
%
N_5G_1^{1(k,l)}\Leftrightarrow N_6G_1^{1(k,l)},\ 
N_7G_1^{1(k,l)}\Leftrightarrow N_8G_1^{1(k,l)}.
\end{eqnarray*}
Thence, again it is sufficient to use the transformations
$N_3$, $N_5$ and $N_7$ in order to generate the first
subgroup of kernels.

The next step refers to the generalisation of the Erd\'elyi
kernels by means of the M\"obius transformations
$K_i$, which now change as well the arguments of the two
hypergeometric functions.  We transform only
the kernel $G_1^{1(1,1)}$ whose explicit
form is
\antiletra
\begin{eqnarray}\label{G-1-lambda}
\textsf{G}_{1}^{1(1,1)}(x,y)&=&\displaystyle
\left(1-\frac{xy}{a}\right) ^{-\lambda}
F\left[ \alpha-\lambda,
\beta-\lambda;\gamma;\frac{xy}{a}\right]
\nonumber\\
&\times&
F\left[\lambda,
\alpha+\beta-\gamma-\lambda;\epsilon;
 \frac{(x-a)(y-a)}{(1-a)(xy-a)}\right].
\end{eqnarray}
The 36 kernels are obtained by replacing each
hypergeometric function
by the other expressions given in Eqs.
(\ref{hiper-1}-c), 
 all of them with the same $\lambda$.

Again we can use only the transformations
$K_{2}$, $K_{3}$, $K_{4}$, $K_{5}$ and
$K_{6}$, the same ones employed in
the  Lambe-Ward case. In effect,
$K_{13}$ gives
\begin{eqnarray*}
K_{13}\textsf{G}_{1}^{1(1,1)}(x,y)&=&\displaystyle
\left(1-\frac{xy}{a}\right) ^{-\lambda}
F\left[ \alpha-\lambda,
\beta-\lambda;\gamma;\frac{xy}{a}\right]
\nonumber\\
&\times&
F\left[\lambda,
\alpha+\beta-\gamma-\lambda;\delta;
 \frac{a(x-1)(y-1)}{(a-1)(xy-a)}\right]=
\textsf{G}_{1}^{1(1,3)}(x,y).
\end{eqnarray*}
Considering the other kernels, we conclude that
$K_{13}$ is equivalent to the identity $K_1$ and,
consequently, the kernels corresponding to the
transformations of every column
of matrix (\ref{matriz-k}) are equivalent to each another.
In addition, as in Lambe-Ward case, we find
$K_i\Leftrightarrow K_{i+6}$ ($i=1,\cdots,6$).

Thus, applying $K_2$,  $K_3$,  $K_4$,  $K_5$
and  $K_6$ on $\textsf{G}_{1}^{1(1,1)}$,
we find  the kernels
\begin{eqnarray}\label{G-2-lambda}
\textsf{G}_{1}^{2(1,1)}(x,y)&=&
\left[ (1-x)(1-y)\right] ^{-\alpha}
\left[1-\frac{(a-1)xy}{a(x-1)(y-1)} \right]^{-\lambda}
\nonumber\\
&\times&
F\left[ \alpha-\lambda,1+\alpha-\delta-\lambda;\gamma;
\frac{(a-1)xy}{a(x-1)(y-1)}\right]\nonumber\\
\nonumber\\
&\times&
F\left[ \lambda,1+2\alpha-\gamma-\delta-\lambda;\epsilon;
\frac{(x-a)(y-a)}{a(1-x-y)+xy}\right],\hspace{1.7cm}
\end{eqnarray}
\begin{eqnarray}
\textsf{G}_{1}^{3(1,1)}(x,y)&=&
\left[ \left(1-\frac{x}{a}\right)\left(1-\frac{y}{a}\right)\right]^{-\alpha}
\left[1-\frac{(1-a)xy}{(x-a)(y-a)} \right]^{-\lambda}
\nonumber\\
\nonumber\\
&\times&
F\left[ \alpha-\lambda,\gamma+\delta-\beta-\lambda;
\gamma;\frac{(1-a)xy}{(a-x)(a-y)}\right]
\nonumber\\
&\times&
F\left[\lambda, \alpha-\beta+\delta-\lambda;\delta;
\frac{(x-1)(y-1)}{a+xy-x-y}
\right],\hspace{2.9cm}
\end{eqnarray}
\begin{eqnarray}
\textsf{G}_{1}^{4(1,1)}(x,y)&=&\left[1- \frac{(x-1)(y-1)}{1-a}\right]^{-\lambda}
F\left[ \alpha-\lambda,
\beta-\lambda;\gamma;\frac{(x-1)(y-1)}{1-a}\right]
\nonumber\\
\nonumber\\
&\times&
F\left[\lambda,\alpha-\beta-\delta-\lambda;\epsilon;
\frac{(x-a)(y-a)}{a(a+xy-x-y)} \right],
\end{eqnarray}
\begin{eqnarray}
\textsf{G}_{1}^{5(1,1)}(x,y)&=&
\left[\left( 1-\frac{x}{a}\right)\left( 1-\frac{y}{a}\right)
\right] ^{-\alpha}
\left[1-\frac{a(x-1)(y-1)}{(x-a)(y-a)} \right]^{-\lambda}
\nonumber\\
\nonumber\\
&\times&
F\left[ \alpha-\lambda,\gamma+\delta-\beta-\lambda;\delta;
\frac{a(x-1)(y-1)}{(x-a)(y-a)}\right]
\nonumber\\
&\times&
F\left[\lambda,\alpha-
\beta+\gamma-\lambda;\gamma;
\frac{xy}{xy-a} \right],\hspace{4.1cm}
\end{eqnarray}
\begin{eqnarray}\label{G-8-lambda}
\textsf{G}_{1}^{6(1,1)}(x,y)&=&\left[1-\frac{(x-a)(y-a)}
{a(a-1)} \right] ^{-\lambda}
F\left[ \alpha-\lambda,\beta-\lambda;\epsilon;
\frac{(x-a)(y-a)}{a(a-1)} \right]
\nonumber\\
\nonumber\\
&\times&
F\left[ \lambda,\gamma+\delta-1-\lambda;\delta;
\frac{a(x-1)(y-1)}{a(1-x-y)+xy} \right] .
\end{eqnarray}
The index transformations suitable for each set are:
$N_2$, $N_5$ and $N_6$ for
 $\textsf{G}_{1}^{2(k,l)}$
and  $\textsf{G}_{1}^{4(k,l)}$;
$N_2$, $N_3$ and $N_4$ for $\textsf{G}_{1}^{3(k,l)}$, $\textsf{G}_{1}^{5(k,l)}$ 
and $\textsf{G}_{1}^{6(k,l)}$.
The previous kernels show that, in addition to the Erd\'elyi substitutions
(\ref{erdelyi-1}) for the independent variables,
there are five other choices for $(\xi,\zeta)$.
\section*{3. Confluent Heun equation}
In this section we  show
that each transformation of the
confluent Heun equation (CHE) is associated with
a transformation of the equation for its kernels. We also
show that some kernels
of Heun's equation lead to kernels for the
CHE through the
limits (\ref{confluence1}).
For this end we rewrite the CHE (\ref{confluence2}) as
\begin{eqnarray}\label{che}
\left[M_{x}-\sigma\right] H(x)=0,\quad
M_{x}=x(x-1)\frac{\partial^2 }{\partial x^2}+
\big[-\gamma+(\gamma+\delta)x+\rho x(x-1)\big]
\frac{\partial}{\partial x}+\alpha\rho x,
\end{eqnarray}
where in the previous equation $M_{x}$ is an ordinary
differential operator.
For the limits (\ref{confluence1}), the integral (\ref{integral-heun})
becomes
\begin{eqnarray}\label{integral-CHE}
\mathcal{H}(x)
=\int_{y_{1}}^{y_{2}}e^{\rho y} \
y^{\gamma-1}(1-y)^{\delta-1}
 \textsf{G}(x,y)H(y)dy,
\end{eqnarray}
where the exponential results from the limit of
$(1-y/a)^{\epsilon-1}$.
Besides this, Eq. (\ref{nucleus}) for $ \textsf{G}(x,y)$
takes the form
\begin{eqnarray}
\label{nucleus-CHE}
\left[M_{x}-{ M}_{y}\right] \textsf{G}(x,y)=0,
\end{eqnarray}
while the expression (\ref{concomitant-heun}) for
the bilinear concomitant becomes
\begin{eqnarray}
\label{concomitant-CHE}
\textsf{P}(x,y)=e^{\rho y}\
y^{\gamma}(1-y)^{\delta}\left[H(y)
\frac{\partial \textsf{G}(x,y)}{\partial y}-\textsf{G}(x,y)\frac{dH(y)}{d y}
\right].
\end{eqnarray}
First we discuss the transformations for the CHE (\ref{che})
and its kernels (\ref{nucleus-CHE}) and, after this,
we explain how to get
kernels for the CHE from the ones of the general
equation (\ref{heun}).
\subsection*{3.1. Transformation of the confluent equation and its kernels}

There are 16 variables
substitutions which preserve the form of
the CHE \cite{decarreaux2}.
If $H(x)=H(\sigma;\rho,\alpha, \gamma,\delta;x)$
denotes one solution of the CHE (\ref{che}),
these transformations are summarised
in the rules $\mathrm{T}_{1},\ \mathrm{T}_{2},\
\mathrm{T}_{3}$ and
$\mathrm{T}_{4}$ that operate as
\begin{eqnarray}
\begin{array}{l}
\mathrm{T}_{1}H(x)=(1-x)^{1-\delta}\
H\big[\sigma-\gamma(1-\delta);\rho,\alpha+1-\delta,
\gamma,2-\delta;x\big],\vspace{2mm}\\
%
\mathrm{T}_{2}H(x)=x^{1-\gamma}\ H
\big[\sigma+(1-\gamma)(\rho-\delta);\rho,
\alpha+1-\gamma,2-\gamma,\delta;x\big],\vspace{2mm}\\
%
\mathrm{T}_{3}H(x)=e^{-\rho x}\ H[\sigma-\gamma\rho;-\rho,
\gamma+\delta-\alpha,
\gamma,\delta;x],\vspace{2mm}\\
%
\mathrm{T}_{4}H(x)=
H\big[\sigma-\rho\alpha;-\rho,\alpha,\delta,\gamma;1-x\big].
\end{array}
\end{eqnarray}
Compositions of these give the group having
16 elements.

On the other side, equation
(\ref{nucleus-CHE}) for the kernels is written in terms
of the same differential operator which appears
in the CHE (\ref{confluence2}). Consequently,
proceeding as in the case of the general Heun equation,
the corresponding rules $ \mathrm{K}_i$
for transforming a given kernel $\textsf{G}(x,y)=
\textsf{G}(\rho,\alpha, \gamma,\delta;x,y))$ are
\begin{eqnarray}\begin{array}{l}
\mathrm{K}_{1}\textsf{G}(x,y)=\big[(1-x)(1-y)\big]^{1-\delta}\
\textsf{G}\big[\rho,\alpha+1-\delta,
\gamma,2-\delta;x,y\big],\vspace{2mm}\\
%
\mathrm{K}_{2}\textsf{G}(x,y)=(xy)^{1-\gamma}\ \textsf{G}
\big[\rho,\alpha+1-\gamma,2-\gamma,\delta;x,y\big],\vspace{2mm}\\
%
\mathrm{K}_{3}\textsf{G}(x,y)=e^{-\rho (x+y)}\ \textsf{G}[ -\rho,
\gamma+\delta-\alpha,
\gamma,\delta;x,y],\vspace{2mm}\\
\mathrm{K}_{4}\textsf{G}(x,y)=
\textsf{G}\big[-\rho,\alpha,\delta,\gamma;1-x,1-y\big].
\end{array}
\end{eqnarray}

These rules can be verified by substitutions of variables.
However, they are useful to produce new kernels
when one knows an initial kernel for the CHE.
In the following we obtain initial kernels
as limits of kernels for the Heun equation.
Any kernel which can be generated by the
above transformations is omitted.

\subsection*{3.2. ``Lambe-Ward-type" kernels}
The limits (\ref{confluence1}) applied to kernels
of Sec. 2.3 give three kinds of kernels
for the CHE, two in terms of confluent
hypergeometric functions and one in terms
of Gauss hypergeometric functions.  The regular and
irregular confluent hypergeometric functions, denoted by $\Phi(\mathrm{a,c};y)$
and $\Psi(\mathrm{a,c};y)$,  respectively, are solutions of the equation
\letra
\begin{eqnarray}\label{confluent0}
u\frac{d^2\varphi}{du^2}+\left( \mathrm{c}-u\right)
\frac{d\varphi}{du} -\mathrm{a}\varphi=0.
\end{eqnarray}
The following types of solutions
for Eq. (\ref{confluent0})
\begin{eqnarray}\label{confluent1}
\begin{array}{ll}
\varphi^{(1)}(u)=\Phi(\mathrm{a,c};u),&\qquad
 \varphi^{(2)}(u)=e^{u}u^{1-\mathrm{c}}\Phi(1-\mathrm{a},2-\mathrm{c};-u),\vspace{2mm}\\
\varphi^{(3)}(u)=\Psi(\mathrm{a,c};u),&\qquad
 \varphi^{(4)}(u)=e^{u}u^{1-\mathrm{c}}\Psi(1-\mathrm{a},2-\mathrm{c};-u),
 \end{array}
\end{eqnarray}
are all defined and distinct only if
$\mathrm{c}$ is not an integer \cite{erdelii}.
Alternative forms for these solutions
follow from the relations
\begin{eqnarray}
\Phi(\mathrm{a,c};u)=e^{u}\Phi(\mathrm{c-a,c};-u),\qquad
\Psi(\mathrm{a,c};u)=u^{1-c}\Psi(\mathrm{1+a-c,2-c};u).
\end{eqnarray}

In the present context, the above confluent hypergeometric
functions result from the limits \cite{erdelii}
\antiletra
\begin{eqnarray}\label{limitCHE}
\begin{array}{l}
\displaystyle\lim_{\mathrm{c}\rightarrow \infty}
F\left(\mathrm{a},\mathrm{b};\mathrm{c};1-\frac{\mathrm{c}}{u}\right)=
\lim_{\mathrm{c}\rightarrow \infty}
F\left(\mathrm{a},\mathrm{b};\mathrm{c};-\frac{\mathrm{c}}{u}\right)=
u^{\mathrm{a}}\Psi(\mathrm{a},\mathrm{a}+1-\mathrm{b};u),\qquad
\vspace{3mm}\\
\displaystyle\lim_{\mathrm{b}\rightarrow \infty}
F\left(\mathrm{a,b;c};\frac{u}{\mathrm{b}}\right)=\Phi(\mathrm
{a,c};u).
\end{array}
\end{eqnarray}
Sometimes, before applying these limits,
it is necessary to use the relations
(\ref{euler-1}) and/or (\ref{euler-2}) and, in addition, multiply
the kernels by suitable constants depending on the
parameter $a$. Furthermore, note that both $a$
and $\beta$ tend to infinity
but such that $\beta= -\rho a$, where $\rho$ is constant. For
this reason we can write, for example,
\begin{eqnarray*}
F\left( \alpha,\beta;\gamma;
\frac{xy}{a}\right)=F\left( \alpha,\beta;\gamma;
\frac{-\rho xy}{\beta}\right),\quad
\left( 1-\frac{xy}{a}\right)^{-\beta}=
\left( 1+\frac{\rho xy}{\beta}\right)^{-\beta}
\end{eqnarray*}
and, thence, accomplish the limit $\beta\to \infty$
by keeping $\rho xy$ fixed.

Thus, the first set of section 2.3,
$\textsf{G}_{1}^{1(i)}(x,y)$, gives the limits
\begin{eqnarray*}\begin{array}{l}
\textsf{G}_{1}^{1(1)}(x,y)\to
\Phi\left( \alpha,\gamma;
-\rho xy\right),
\vspace{2mm}\\
%
\textsf{G}_{1}^{1(2)}(x,y)\to
e^{-\rho xy}
\left( {xy}\right)^{1-\gamma}
\Phi\left( 1-\alpha,
2-\gamma;\rho xy\right),\vspace{2mm}\\
%
{\textsf{G}}_{1}^{1(3)}(x,y)\text{ and } {\textsf{G}}_{1}^{1(5)}(x,y)\to
\Psi\left( \alpha,\gamma;-\rho xy\right),\vspace{2mm}\\
%
{\textsf{G}}_{1}^{1(4)}(x,y) \text{ and }
{\textsf{G}}_{1}^{1(6)}(x,y)\to e^{-\rho xy}(xy)^{1-\gamma}
\ \Psi\left(1- \alpha,2-\gamma;
\rho xy\right),
\end{array}
\end{eqnarray*}
which, by means of (\ref{confluent1}), can be written as
\begin{eqnarray}
G_1^{(i)}(x,y)=\varphi^{(i)}(u),\ \text{ with}\quad u=-\rho xy,
\quad \mathrm{a}=\alpha,\quad
\mathrm{c}=\gamma\quad [i=1,2,3,4].
\end{eqnarray}
This initial set of kernels is really due to Lambe and
Ward because it is given by confluent hypergeometric
functions whose arguments depend
on the product $xy$ as in \cite{lambe}.
The previous transformations $\mathrm{K}_i$ produce
new solutions of this type.

The limits of the set $\textsf{G}_{1}^{2(i)}(x,y)$, Sec. 2.3,
gives six kernels
in terms of Gauss hypergeometric
functions (\ref{hiper-1}-c), written as
\begin{eqnarray}
&&\tilde{G}_{1}^{(i)}(x,y)=
\left[ (1-x)(1-y)\right] ^{-\alpha}F^{(i)}(u), \qquad [i=1,\cdots,6]\\
\nonumber\\
&&\text{ with}\quad u=\frac{xy}{(x-1)(y-1)},
\quad \mathrm{a}=\alpha,\quad \mathrm{b}=1+\alpha-\delta,\quad \mathrm{a}=\gamma.\nonumber
\end{eqnarray}
We can also apply the transformations
$\mathrm{K}_j$ in order to generate a group of kernels. Notice that this
group results from a generalisation of the Lambe-Ward kernels and,
as far as we know, it is new.

Finally, another set of kernels is given by
confluent hypergeometric functions (\ref{confluent1})
whose arguments are proportional to $x+y-1$. It is obtained
as limits of the kernels (\ref{lambe-8}) and reads
\begin{eqnarray}
\hat{G}_1^{(i)}=\varphi^{(i)}(u),\quad \text{with} \ u=-\rho (x+y-1),
\quad \mathrm{a}=\alpha,\quad\mathrm{c}=\gamma+\delta \quad [i=1,2,3,4].
\end{eqnarray}
Other kernels follow from the transformations $K_i$.
This group is also a result of a generalisation
of the Lambe-Ward kernels, but kernels having this form
are already known in the literature \cite{slavyanov,ronveaux}.

\subsection*{3.3. ``Erd\'elyi-type" kernels}

By taking the limits of the Erd\'elyi-type kernels we find
two groups of kernels for the CHE. The
group given by products of confluent hypergeometric
functions has already appeared in the literature \cite{masuda},
whereas the group given by products of hypergeometric
and confluent hypergeometric functions seems to be new.
In the limit process we suppose that
$\lambda$ is kept fixed,
that is, we assume that $\lambda$
does not depend on the parameters $\beta$ and $a$ of
the Heun equation.

Thus, by taking the limit of the kernels $\textsf{G}_{1}^{1(k,l)}(x,y)$
given in (\ref{erdelyi-2}), we
get an initial set of kernels given by
\begin{eqnarray}
G_{1}^{(i,j)}=
\varphi^{(i)}(\xi)\bar{\varphi}^{(j)}(\zeta),\qquad [i,j=1,2,3,4]
\end{eqnarray}
where $\varphi^{(i)}(\xi)$ and $\bar{\varphi}^{(j)}(\zeta)$ are
the solutions (\ref{confluent1}) for the confluent hypergeometric
equation, having the following
arguments and parameters :
\begin{eqnarray}
 \displaystyle
\varphi^{i}(\xi):\quad \xi=-\rho xy, \
\mathrm{a}=\alpha-\lambda,\
\mathrm{c}=\gamma;\quad
 \displaystyle
\bar{\varphi}^{j}(\zeta):\quad \zeta=\rho(x-1)(y-1), \
\mathrm{a}= \lambda,\
\mathrm{c}=\delta.
\end{eqnarray}
The four kernels given by products of regular functions
$\Phi$ are
\begin{eqnarray*}
G_1^{(1,1)}(x,y)=\Phi\left[\alpha-\lambda,\gamma;-\rho xy \right]
\Phi\left[\lambda,\delta;\rho(x-1)(y-1) \right],
\hspace{3cm}
\end{eqnarray*}
\begin{eqnarray*}
G_1^{(1,2)}(x,y)&=&e^{\rho(x-1)(y-1)}
\left[ (x-1)(y-1)\right]^{1-\delta}
\Phi\left[\alpha-\lambda,\gamma;-\rho xy \right]\hspace{2.1cm}
\\
\\
&\times&
\Phi\left[1-\lambda,2-\delta;-\rho(x-1)(y-1) \right],
\end{eqnarray*}
\begin{eqnarray*}
G_1^{(2,1)}(x,y)=e^{-\rho xy}(xy)^{1-\gamma}
\Phi\left[\lambda+1-\alpha,2-\gamma;\rho xy \right]
\Phi\left[\lambda,\delta;\rho(x-1)(y-1) \right],
\end{eqnarray*}
\begin{eqnarray*}
G_1^{(2,2)}(x,y)&=&e^{-\rho(x+y)}(xy)^{1-\gamma}
\left[(x-1)(y-1) \right]^{1-\delta}
\Phi\left[\lambda+1-\alpha,2-\gamma;\rho xy \right]\hspace{.5cm}
\\
\\
&\times&
\Phi\left[1-\lambda,2-\delta;-\rho(x-1)(y-1) \right].
\end{eqnarray*}
Replacing one or both $\Phi$ by $\Psi$ we get
the set with 16 kernels. The other sets
result from this by the transformations $\mathrm{K}_i$.

In the second place, the kernel $\textsf{G}_{1}^{6(1,1)}(x,y)$,
Eq. (\ref{G-8-lambda}), yields a kernel $\tilde{G}_{1}^{(2,1)}(x,y)$ 
constituted by a product of hypergeometric and
confluent hypergeometric functions, namely,
\begin{eqnarray*}
\tilde{G}_{1}^{(2,1)}(x,y)&=&\left(1-x-y\right)^{-\lambda}
\Psi\left[\alpha-\lambda,\gamma+\delta-2\lambda;
\rho (1-x-y)\right]
\nonumber\\
\nonumber\\
&\times&
F\left[\lambda,\gamma+\delta-1-\lambda;\delta;
\frac{(x-1)(y-1)}{1-x-y} \right] .
\end{eqnarray*}
By considering the limits of the full set $\textsf{G}_{1}^{6(k,l)}(x,y)$
we obtain the initial set
\begin{eqnarray}
\tilde{G}_{1}^{(i,j)}(x,y)=\varphi^{(i)}(\xi)F^{(j)}(\zeta),\qquad [i=1,\cdots,4;\ j=1,\cdots,6]
\end{eqnarray}
where $\varphi^{(i)}$ are the four
solutions for confluent hypergeometric equation and where $F^{(j)}$
are the six solutions for hypergeometric equation
with the following arguments and parameters:
\begin{eqnarray}\label{quinto-grupo-b}
\begin{array}{lllll}
 \displaystyle
\varphi^{(i)}(\xi):&\quad \xi=\rho (1-x-y),& \quad
\mathrm{a}=\alpha-\lambda,& \quad \mbox{}&\quad
\mathrm{c}=\gamma+\delta-2\lambda\vspace{3mm}\\
 \displaystyle
F^{(j)}(\zeta):&\quad \zeta=\frac{(x-1)(y-1)}{1-x-y},& \quad
\mathrm{a}= \lambda,&\quad \mathrm{b}=\gamma+\delta-1-\lambda, &\quad
\mathrm{c}=\delta.
    \end{array}
\end{eqnarray}
The kernels $\textsf{G}_{1}^{2(k,l)}(x,y)$, Eq.(\ref{G-2-lambda}),
also lead to kernels given by products hypergeometric and
confluent hypergeometric functions, but we may show that
these are connected with the previous ones by the transformation
$\mathrm{K}_4$.

\section*{4. Concluding remarks}

We have taken the following steps to deal with kernels for integral relations
among solutions of Heun equations:
\begin{itemize}
\itemsep-3pt
   \item
the use of an integral with a weight function
$w(x,y)$ which allows to write a given Heun equation
and the respective equation for the kernels  in 
terms of operators functionally identical, say,
$M_x$ and $M_y$;
\item the use of the known transformations
of the equation in order to get the actual form 
for the transformations of the kernel equation;
\item the generation of new kernels by applying the previous transformations
to an initial kernel or set of kernels.
\item the use of a limiting procedure to generate 
kernels for the confluent Heun equation.
\end{itemize}
For the (general) Heun equation
we have used Maier's transformations, 
discarding as inappropriate the forms given 
in Refs. \cite{arscott2} and \cite{nist} for the 
homotopic transformations. As initial
kernels we have employed the ones found by Lambe and Ward, Eq. (\ref{def-Lambe}), and
by Erd\'elyi, Eq. (\ref{erdelyi-2}).

In this manner, in section 2 the transformations for integral relations have
afforded several new kernels for the Heun equation, given by a single
hypergeometric function
and by products of two hypergeometric functions. 
We have seen that only six of the homographic 
transformations for the kernels are effective, namely: $K_1$, $K_2$,$\cdots$,$K_6$,
where $K_1$ is the identity. The fact these are just the six first
transformations of Appendix B is a consequence of manner in which 
we have written the elements of matrix (\ref{24fracional}).
 
We have written only some of the possible kernels,
but a wider list can be generated by index transformations
which lead to new kernels
where the hypergeometric functions possess
the same argument but different parameters.
In addition, from a kernel with a given argument,
new kernels follow from the fact that the hypergeometric
equation formally admits solutions
with different arguments in the vicinity of each singular point.

In section 3, the confluence
procedure (\ref{confluence1}) has led to 
five sets of initial kernels for the CHE, three of
them arising from generalisations
of the Lambe-Ward and Erd\'elyi kernels by means of 
M\"obius transformations. This in association with the Leaver version
for the CHE and the concept of Whittaker-Ince limit suggest
new kernels also for the double-confluent Heun equation (DCHE)
and for limiting cases of the CHE and DCHE.
In effect, by substitutions of variables, the
CHE (\ref{confluence2}) can be written
in the Leaver form \cite{leaver}, namely,
\begin{eqnarray}
\label{gswe}
z(z-z_{0})\frac{d^{2}U}{dz^{2}}+(B_{1}+B_{2}z)
\frac{dU}{dz}+
\left[B_{3}-2\eta
\omega(z-z_{0})+\omega^{2}z(z-z_{0})\right]U=0,
\quad [\omega\neq0]
\end{eqnarray}
where $B_{i}$, $\eta$ and $\omega$ are constants,
and $z=0$ and $z=z_{0}$ are the regular singular
points. When $z_{0}=0$ (Leaver's limit), this gives the DCHE
\begin{eqnarray}\label{dche-}
z^{2}\frac{d^{2}U}{dz^{2}}+
\left(B_{1}+B_{2}z\right)\frac{dU}{dz}+
\left(B_{3}-2\eta \omega z+\omega^{2}z^{2}\right)U=0,
\qquad \left[B_{1}\neq 0, \ \omega\neq 0\right],
\end{eqnarray}
where now $z=0$ and $z=\infty$ are irregular
singularities. On the other side, these equations admit the
limit \cite{eu}
\begin{eqnarray*}\label{ince}
\omega\rightarrow 0, \ \
\eta\rightarrow
\infty, \ \mbox{such that }\ Â \ 2\eta \omega =-q,\qquad
[\mbox{Whittaker-Ince limit}]
\end{eqnarray*}
where $q$ should not be confused with
the parameter $q$ of the Heun equation (\ref{heun}).
The Whittaker-Ince limit of the CHE and DCHE are, respectively,
the equations
\begin{eqnarray}
\label{incegswe}
&&z(z-z_{0})\frac{d^{2}U}{dz^{2}}+(B_{1}+B_{2}z)
\frac{dU}{dz}+
\left[B_{3}+q(z-z_{0})\right]U=0,\qquad [q\neq0]\\
&&z^2\frac{d^{2}U}{dz^{2}}+(B_{1}+B_{2}z)
\frac{dU}{dz}+
\left(B_{3}+qz\right)U=0,\qquad [q\neq0, \ B_{1}\neq 0]
\end{eqnarray}
which have a different type of singularity at $x=\infty$
as compared with the original CHE and DCHE \cite{eu,lea,lea-2}.
As the preceding limits connect the CHE with the DCHE and
their respective Whittaker-Ince limits,
we may expect to find kernels for each of these
out of kernels arising from the Heun equation.
Thus, by developing the results of section 3 
we could unify the treatment of these equations.
We advance that we will find that the usual kernels for the
Mathieu equation \cite{McLachlan} are particular 
cases of the kernels for Eq. (\ref{incegswe}).

The construction of new kernels can be envisaged as a first step for
seeking new solutions for
the Heun equations by means of integral relations \cite{erdelyi4}. 
However, as in the case of the Mathieu 
equation \cite{McLachlan}, it is not easy to use this technique. 
Thus, in the following we focus on possibility of using some of the 
Maier transformations to extend the solutions in series of hypergeometric 
functions given by Svartholm in 1939 \cite{svartholm} and by Erd\'elyi 
in 1944 \cite{erdelyi3}. Also in this context, the previous limits 
become relevant.

As a prior consideration, we note that there are 
three types of recurrence relations for the series coefficients -- not
just one as given in the original articles and repeated since then \cite{arscott2,nist}.
The two additional relations may be found 
by the procedure used in Appendix A of Ref. \cite{eu2}. Then,
by virtue of the new relations and by
means of the homotopic transformations, 
we may show that the Svartholm solutions include as particular cases 
the eight Fourier-type solutions found by Ince in 1940 
for the Lam\'e equation \cite{nist,ince3}.

On the other side, both the Svartholm and the Erd\'elyi 
solutions are valid only if the parameter $a$ satisfy the condition
 $\mathrm{Re}\sqrt{1-(1/a)}>0$ which is assured by requiring that $a\not\in[0,1]$. 
In addition, the former solution is given by a single expansion
in series of hypergeometric functions and converges only over
a finite region of the complex plane; however, the latter is in fact
a set of expansions in terms of hypergeometric functions
which, by analytical continuation, may cover the entire complex plane
provided that a characteristic equation is fulfilled.

Then, Erd\'elyi's
solutions are candidates to solve a cosmological problem formulated  
by Kantowski \cite{kantowski} because in this case the variable $x$ 
extend to infinity. Nevertheless, the problem 
demands solutions valid also for $a\in[0,1]$. These
may be derived by applying on the Erd\'elyi
solutions one of the following linear transformations:
\begin{eqnarray*}
\displaystyle
&M_{9}H(x)=H\left(\frac{1}{a},\frac{q}{a};\alpha,\beta,\gamma,\epsilon;\frac{x}{a}\right),\quad
M_{61}H(x)=H\left(\frac{1}{1-a},\frac{q-\alpha\beta}{a-1};\alpha,\beta,\delta,\epsilon;\frac{x-1}{a-1}\right),
\\
\displaystyle
&M_{101}H(x)=H\left(\frac{a-1}{a},\frac{-q+\alpha\beta a}{a};\alpha,\beta,\epsilon,\gamma;\frac{a-x}{a}\right),
\\
&M_{109}H(x)=H\left(\frac{a}{a-1},\frac{-q+\alpha\beta a}{a-1};\alpha,\beta,\epsilon,\delta;\frac{a-x}{a-1}\right).
\end{eqnarray*}
Hence we find, respectively, the conditions:
$a\not\in[1,\infty)$, $a\not\in(-\infty,0]$, $a\not\in$$[1,\infty)$
and $a\not\in(-\infty,0]$. However, for each case it is necessary to 
reexamine the domains of convergence.

After these preliminaries, 
we conclude by adding that
some of the Svartholm and Erd\'elyi solutions
-- the solution (\ref{erdelyi-1944}), for example -- lead to solutions for the
CHE by means of the limits (\ref{confluence1}). To
prove this it is sufficient to divide the recurrence
relations by the parameter $a$ before performing the limits. 
Furthermore, we can show as well that the solutions for the CHE in the form (\ref{gswe}) 
supply solutions for the DCHE (\ref{dche-}) through the 
Leaver limit ($z_0\to 0$). These are additional reasons   
for choosing the Svartholm and Erd\'elyi solutions
as a starting point for further investigation.

%
%
%
\section*{Appendix A. Equations of the first section}
\protect\label{A}
\setcounter{equation}{0}
\renewcommand{\theequation}{A\arabic{equation}}
Equations (\ref{nucleus})
and (\ref{concomitant-heun}) as well as the
condition $\textsf{P}(x,y_1)=\textsf{P}(x,y_2)$
are obtained from the general
theory of integral relations \cite{ince1} which is
established for $w(x,y)=1$,  that is, for
\begin{eqnarray}\label{integral-heun-2}
\mathcal{H}(x)=\int_{y_{1}}^{y_{2}}\mathbb{K}(x,y)H(y)dy
\end{eqnarray}
where $\mathbb{K}(x,y)$ denotes the kernel. In this case
the equation for $\mathbb{K}(x,y)$ is given in terms of
the operators ${M}_{x}$ and $\bar{M}_{y}$, where
$\bar{M}_{y}$ is the adjoint operator \cite{ince1}
corresponding to ${M}_{y}$, that is,
\begin{eqnarray*}
\bar{M}_{y}&=&y(y-1)(y-a)
\frac{\partial^{2}}{\partial y^2}+
\big[(2-\gamma)(y-1)(y-a)+(2-\delta){y}(y-a)+
(2-\epsilon){y}(y-1)\big]\frac{\partial}{\partial y}\nonumber\\
&+&\big[4-2(\alpha+\beta+1)+\alpha \beta \big]y+
a(\gamma+\delta-2)+\epsilon+\gamma-2.
\end{eqnarray*}
By applying $M_{x}$ to the integral (\ref{integral-heun-2})

and supposing that the integration endpoints are
independent of $x$, we find
\begin{eqnarray*}
M_{x}\mathcal{H}(x)=\int_{y_{1}}^{y_{2}}H(y)
\left[ M_{x}-\bar{M}_{y}\right] \mathbb{K}(x,y)dy+
\int_{y_{1}}^{y_{2}} H (y)\bar{M}_{y}\mathbb{K}(x,y)dy.
\end{eqnarray*}
Then, by requiring that the kernel satisfies
the partial differential
equation
\begin{eqnarray}\label{nucleus-2}
\left[M_{x}-\bar{ M}_{y}\right] \mathbb{K}(x,y)=0,
\end{eqnarray}
the right side of of the previous integral
reduces to $\int H(y)\bar{M}_{y}\mathbb{K}(x,y)dy$. Using
the Lagrange identity
\begin{eqnarray*}
H(y)\bar{M}_{y}\mathbb{K}(x,y)=\mathbb{K}(x,y) M_{y}H(y)+
\frac{\partial}{\partial y}\textsf{P}(x,y)\stackrel{\mbox{(\ref{heun-2})}}=
q\mathbb{K}(x,y) H(y)+
\frac{\partial}{\partial y}\textsf{P}(x,y),
\end{eqnarray*}
where now $\textsf{P}(x,y)$
is given by
\begin{eqnarray}
\label{concomitant-heun-ince}
\textsf{P}(x,y)&=&
y(y-1)(y-a)\left[H(y)
\frac{\partial \mathbb{K}(x,y)}{\partial y}-\mathbb{K}(x,y)\frac{dH(y)}{d y}
\right]\nonumber\\
&+&\Big[(1-\gamma)(y-1)(y-a)+(1-\delta)y(y-a)+
(1-\epsilon)y(y-1)\Big]H(y)\mathbb{K}(x,y),
\end{eqnarray}
then we find that the integral yields
\begin{eqnarray}
M_{x}\mathcal{H}(x)=
\int_{y_{1}}^{y_{2}}\left[q\mathbb{K}(x,y)H(y)+
\frac{\partial{\textsf{P}(x,y)}}{\partial{y}}\right]dy 
\stackrel{\mbox{(\ref{integral-heun-2})}}\Longleftrightarrow
[M_{x}-q]\mathcal{H}(x)=
\textsf{P}(x,y)\Big|_{y=y_1}^{y=y_2}.
\end{eqnarray}
Therefore, $\mathcal{H}(x)$
will be a solution of the Heun equation
if $\mathbb{K}$ is solution of (\ref{nucleus-2}),
if the integral (\ref{integral-heun-2})
exists and the limits of integration are so chosen that
$\textsf{P}(x,y_1)=\textsf{P}(x,y_2)$.
Further, by setting $\mathbb{K}(x,y)=w(x,y)\textsf{G}(x,y)$, 
where $w(x,y)$ is defined in Eq. (\ref{integral-heun}), 
we  recover equations (\ref{nucleus}) and
(\ref{concomitant-heun}). Notice that Eq.
(\ref{nucleus-2}) is  inadequate to deal with
the kernel transformations because the operators
$M_{x}$ and $\bar{ M}_{y}$ do not present
the same functional form.

%
%
\section*{Appendix B. M\"obius transformations for kernels}
\protect\label{B}
\setcounter{equation}{0}
\renewcommand{\theequation}{B\arabic{equation}}
As in the index transformations,
firstly we find the M\"obius transformations
$M_i$ for $H(x)$
in Maier's table and, then, write the kernel
transformation $K_i$ in accordance with
the rule (\ref{fraction-G}). 
Thus, the 24 expressions
for the $K_i$ of matrix (\ref{matriz-k}) are the ones given below.
\begin{eqnarray*}
K_1\textsf{G}(x,y)=\textsf{G}(x,y)=
\textsf{G}\left[a; \alpha,\beta,\gamma,\delta;x,y\right],
\qquad[\text{Identity}].\hspace{4.1cm}
\end{eqnarray*}
\begin{eqnarray*}
K_2\textsf{G}(x,y)=  \left[ (1-x)(1-y)\right] ^{-\alpha}
\textsf{G}\left[\frac{a}{a-1}; \alpha,1+\alpha-\delta,\gamma,1
+\alpha-\beta;\frac{x}{x-1},
\frac{y}{y-1}\right].\ \
\end{eqnarray*}
\begin{equation*}
K_{3}\textsf{G}(x,y)= \left[ \left(1-\frac{x}{a}\right)
\left(1-\frac{y}{a}\right)\right] ^{-\alpha}\textsf{G}\left[\frac{1}{1-a}; \alpha,\gamma+\delta-\beta,\gamma,1+\alpha-\beta;\frac{x}{x-a},
\frac{y}{y-a}\right].
\end{equation*}
\begin{eqnarray*}
K_{4}\textsf{G}(x,y)= \displaystyle
\textsf{G}\left[ {1}-{a};\alpha, \beta,\delta,\gamma;
1-x,1-y\right].\hspace{6cm}
\end{eqnarray*}
\begin{eqnarray*}
K_{5}\textsf{G}(x,y)=\displaystyle
\left[\left(1- \frac{x}{a}\right)\left( 1-\frac{y}{a}\right)
\right] ^{-\alpha}
\textsf{G}\left[ \frac{1}{a};\alpha, \gamma+\delta-\beta,
\delta,1+\alpha-\beta;
\frac{x-1}{x-a},\frac{y-1}{y-a}\right].\hspace{.2cm}
\end{eqnarray*}
\begin{eqnarray*}
K_{6}\textsf{G}(x,y)= \displaystyle
\textsf{G}\left[ \frac{a-1}{a};\alpha,
\beta,\epsilon,\gamma;
\frac{a-x}{a},\frac{a-y}{a}\right].\hspace{5.6cm}
\end{eqnarray*}
\begin{eqnarray*}
K_{7}\textsf{G}(x,y)=
\left( xy\right)^{-\alpha} \displaystyle
\textsf{G}\left[ \frac{1}{a};\alpha, 1+\alpha-\gamma,
1+\alpha-\beta,\delta;
\frac{1}{x},\frac{1}{y}\right].\hspace{3.7cm}
\end{eqnarray*}
\begin{eqnarray*}
K_{8}\textsf{G}(x,y)=
\left( xy\right)^{-\alpha} \displaystyle
\textsf{G}\left[ \frac{a-1}{a};\alpha, 1+\alpha-\gamma,
\delta,1+\alpha-\beta;
\frac{x-1}{x},\frac{y-1}{y}\right].\hspace{1.9cm}
\end{eqnarray*}
\begin{eqnarray*}
 K_{9}\textsf{G}(x,y)=
\left( {xy} \right) ^{-\alpha}
\displaystyle
\textsf{G}\left[ 1-a;\alpha, 1+\alpha-\gamma,
\epsilon,1+\alpha-\beta;
\frac{x-a}{x},\frac{y-a}{y}\right].\hspace{2cm}
\end{eqnarray*}
\begin{eqnarray*}
K_{10}\textsf{G}(x,y)=
\left[ (1-x)(1-y)\right] ^{-\alpha}\textsf{G}\left[\frac{1}{1-a}; \alpha,1+\alpha-\delta,1+\alpha-\beta,\gamma;\frac{1}{1-x},
\frac{1}{1-y}\right].\
\end{eqnarray*}
\begin{eqnarray*}
K_{11}\textsf{G}(x,y)=
{\left[\left( {1-x}\right)\left( {1-y}\right)
\right] ^{-\alpha}}
\displaystyle
\textsf{G}\left[ a;\alpha, 1+\alpha-\delta,
\epsilon,1+\alpha-\beta;
\frac{x-a}{x-1},\frac{y-a}{y-1}\right].\hspace{1cm}
\end{eqnarray*}
\begin{equation*}
K_{12}\textsf{G}(x,y)=
\left[\left( 1-\frac{x}{a}\right)\left( 1-\frac{y}{a}\right)
\right] ^{-\alpha}
\textsf{G}\left[ \frac{a}{a-1};\alpha, \gamma+\delta-\beta,
1+\alpha-\beta,\gamma;
\frac{a}{a-x},\frac{a}{a-y}\right].
\end{equation*}
\begin{eqnarray*}
K_{13}\textsf{G}(x,y)=
\textsf{G}\left[\frac{1}{a}; \alpha,\beta,\gamma,\epsilon;\frac{x}{a},
\frac{y}{a}\right].\hspace{7.4cm}
\end{eqnarray*}
\begin{eqnarray*}
K_{14}\textsf{G}(x,y)=  \left[ (1-x)(1-y)\right] ^{-\alpha}
\textsf{G}\left[\frac{a-1}{a}; \alpha,1+\alpha-\delta,\gamma,\epsilon;
\frac{(a-1)x}{a(x-1)},
\frac{(a-1)y}{a(y-1)}\right].\hspace{.5cm}
\end{eqnarray*}
\begin{equation*}
K_{15}\textsf{G}(x,y)= \left[ \left(1-\frac{x}{a}\right)
\left(1-\frac{y}{a}\right)\right] ^{-\alpha}
\textsf{G}\left[{1-a}; \alpha,\gamma+\delta-\beta,\gamma,\delta;
\frac{(1-a)x}{x-a},
\frac{(1-a)y}{y-a}\right].\hspace{.3cm}
\end{equation*}
\begin{eqnarray*}
K_{16}\textsf{G}(x,y)= \displaystyle
\textsf{G}\left[ \frac{1}{1-a};\alpha, \beta,\delta,\epsilon;
\frac{1-x}{1-a},\frac{1-y}{1-a}\right].\hspace{5.5cm}
\end{eqnarray*}
\begin{eqnarray*}
K_{17}\textsf{G}(x,y)=\displaystyle
\left[\left( 1-\frac{x}{a}\right)\left(1- \frac{y}{a}\right)
\right] ^{-\alpha}
\textsf{G}\left[ {a};\alpha, \gamma+\delta-\beta,
\delta,\gamma;
\frac{a(x-1)}{x-a},\frac{a(y-1)}{y-a}\right].\hspace{.5cm}
\end{eqnarray*}
\begin{eqnarray*}
K_{18}\textsf{G}(x,y)= \displaystyle
\textsf{G}\left[ \frac{a}{a-1};\alpha,
\beta,\epsilon,\delta;
\frac{a-x}{a-1},\frac{a-y}{a-1}\right].\hspace{5.45cm}
\end{eqnarray*}
\begin{eqnarray*}
K_{19}\textsf{G}(x,y)=
\left( xy\right)^{-\alpha} \displaystyle
\textsf{G}\left[ {a};\alpha, 1+\alpha-\gamma,
1+\alpha-\beta,{\epsilon};
\frac{a}{x},\frac{a}{y}\right].\hspace{3.7cm}
\end{eqnarray*}
\begin{eqnarray*}
K_{20}\textsf{G}(x,y)=
\left( xy\right)^{-\alpha} \displaystyle
\textsf{G}\left[ \frac{a}{a-1};\alpha, 1+\alpha-\gamma,
\delta,\epsilon;
\frac{a(x-1)}{(a-1)x},\frac{a(y-1)}{(a-1)y}\right].\hspace{2.2cm}
\end{eqnarray*}
\begin{eqnarray*}
 K_{21}\textsf{G}(x,y)=
\left( {xy} \right) ^{-\alpha}
\displaystyle
\textsf{G}\left[ \frac{1}{1-a};\alpha, 1+\alpha-\gamma,
\epsilon,\delta;
\frac{x-a}{{(1-a)}x},
\frac{y-a}{{(1-a)}y}\right].\hspace{2.2cm}
\end{eqnarray*}
\begin{eqnarray*}
K_{22}\textsf{G}(x,y)=
\left[ (1-x)(1-y)\right] ^{-\alpha}\textsf{G}\left[{1-a}; \alpha,1+\alpha-\delta,1+\alpha-\beta,\epsilon;\frac{1-a}{1-x},
\frac{1-a}{1-y}\right].\quad
\end{eqnarray*}
\begin{eqnarray*}
K_{23}\textsf{G}(x,y)=
{\left[\left({1-x}\right)\left( {1-y}\right)
\right] ^{-\alpha}}
\displaystyle
\textsf{G}\left[ \frac{1}{a};\alpha, 1+\alpha-\delta,
\epsilon,\gamma;
\frac{x-a}{a(x-1)},\frac{y-a}{a(y-1)}\right].\hspace{1.2cm}
\end{eqnarray*}
\begin{equation*}
K_{24}\textsf{G}(x,y)=
\left[\left( 1-\frac{x}{a}\right)\left( 1-\frac{y}{a}\right)
\right] ^{-\alpha}
\textsf{G}\left[ \frac{a-1}{a};\alpha, \gamma+\delta-\beta,
1+\alpha-\beta,\delta;
\frac{1-a}{{x-a}},
\frac{1-a}{{y-a}}\right].
\end{equation*}
%

%
%

%
\end{document}